\documentclass[12pt]{amsart}

\usepackage{amssymb,amsmath,amsaddr}
\usepackage{bm}
\usepackage{graphicx,hyperref}

\newtheorem{theo}{Theorem}[section]
\newtheorem{lemm}[theo]{Lemma}
\theoremstyle{remark}
\newtheorem{example}[theo]{Example}

\numberwithin{equation}{section}

\begin{document}

\hfill{Dedicated to the 60th birthday of Iskander Taimanov} 

\vspace{1cm}

\title{Geometry of quasiperiodic functions on the plane}

\author{I.A. Dynnikov$^{1}$, A.Ya. Maltsev$^{2}$, S.P. Novikov$^{1,2}$}

\address{
\centerline{$^{1}$ \it{V.A. Steklov Mathematical Institute 
of Russian Academy of Sciences}}
\centerline{\it 119991 Moscow, Gubkina str. 8}
\centerline{$^{2}$ \it{L.D. Landau Institute for Theoretical Physics 
of Russian Academy of Sciences}}
\centerline{\it 142432 Chernogolovka, pr. Ak. Semenova 1A}
}

\begin{abstract}
The present article proposes a review of the most recent results obtained in the study 
of Novikov's problem on the description of the geometry of the level lines
 of quasi-periodic functions in the plane. Most of the paper is devoted to the 
results obtained for functions with three quasi-periods, which play a very 
important role in the theory of transport phenomena in metals. In this part, along 
with previously known results, a number of new results are presented that 
significantly refine the general description of the picture that arises in this case.
New statements are also presented for the case of functions with more than three 
quasi-periods, which open up approaches to the further study of Novikov's problem 
in the most general formulation. The role of Novikov's problem in various fields 
of mathematical and theoretical physics is also discussed.
\end{abstract}

\maketitle

\section{Introduction}

 The theory of quasi-periodic functions originates in the works of G. Bohr and 
A.S. Besikovich (\cite{Bohr,Besicovitch}) and plays an important role in the description of
a huge variety of phenomena in various fields of theoretical and applied science. 
As a rule, a quasi-periodic function $f (x^{1}, \dots, x^{n})$ with $N$ quasi-periods in the space 
$\mathbb{R}^{n}$ is the restriction of a ``good enough'' 
(for example, smooth) $N$-periodic function $F (z^{1}, \dots, z^{N})$ onto the image 
of the space $\mathbb R^n$ under some affine embedding 
$\iota:\mathbb{R}^{n} \hookrightarrow \mathbb{R}^{N}$, i.e. $f=F\circ\iota$.
In the case of a generic embedding~$\iota$, the function $f (x^{1}, \dots, x^{n})$ 
has no exact periods in $\mathbb{R}^{n}$. Such periods appear when the intersection 
of the subspace~$\iota(\mathbb R^n)$ with the lattice~$\mathbb Z^N$ is not empty.
It may also turn out that the image~$\iota(\mathbb R^n)$ is contained in a non-trivial 
subspace of an integral direction. In this case, the number of quasi-periods of the 
function~$f$ will be less than~$N$.

 In this paper, both the generic cases and special cases of quasi-periodic functions 
corresponding to embeddings $\iota: \mathbb{R}^{n} \hookrightarrow \mathbb{R}^{N}$ 
of various types will be important for us.

 It is well known that the theory of quasiperiodic functions is extremely important 
in the description of quasicrystals. In this case, the physically important cases 
are $n = 2$ and $n = 3$, and $N$ is most often equal to $2n$. In addition, the theory 
of quasi-periodic functions underlies the description of solutions to integrable 
dynamical systems (both finite- and infinite-dimensional).

 In this survey, we consider qualitative questions of the geometry of quasiperiodic 
functions on the plane. By this we mean the global behavior of their level lines 
$f (x, y) = {\rm const}$, which plays a very important role in the description of a large 
number of physical phenomena. The problem of a qualitative description of the 
geometry of the level lines of quasi-periodic functions on the plane 
(Novikov's problem) is very non-trivial, and its complexity grows rapidly 
with an increase in the number of quasi-periods. Here we will try to give an 
overview of the most recent results obtained in this area.

\begin{figure}[t]
\begin{center}
\includegraphics[width=0.9\linewidth]{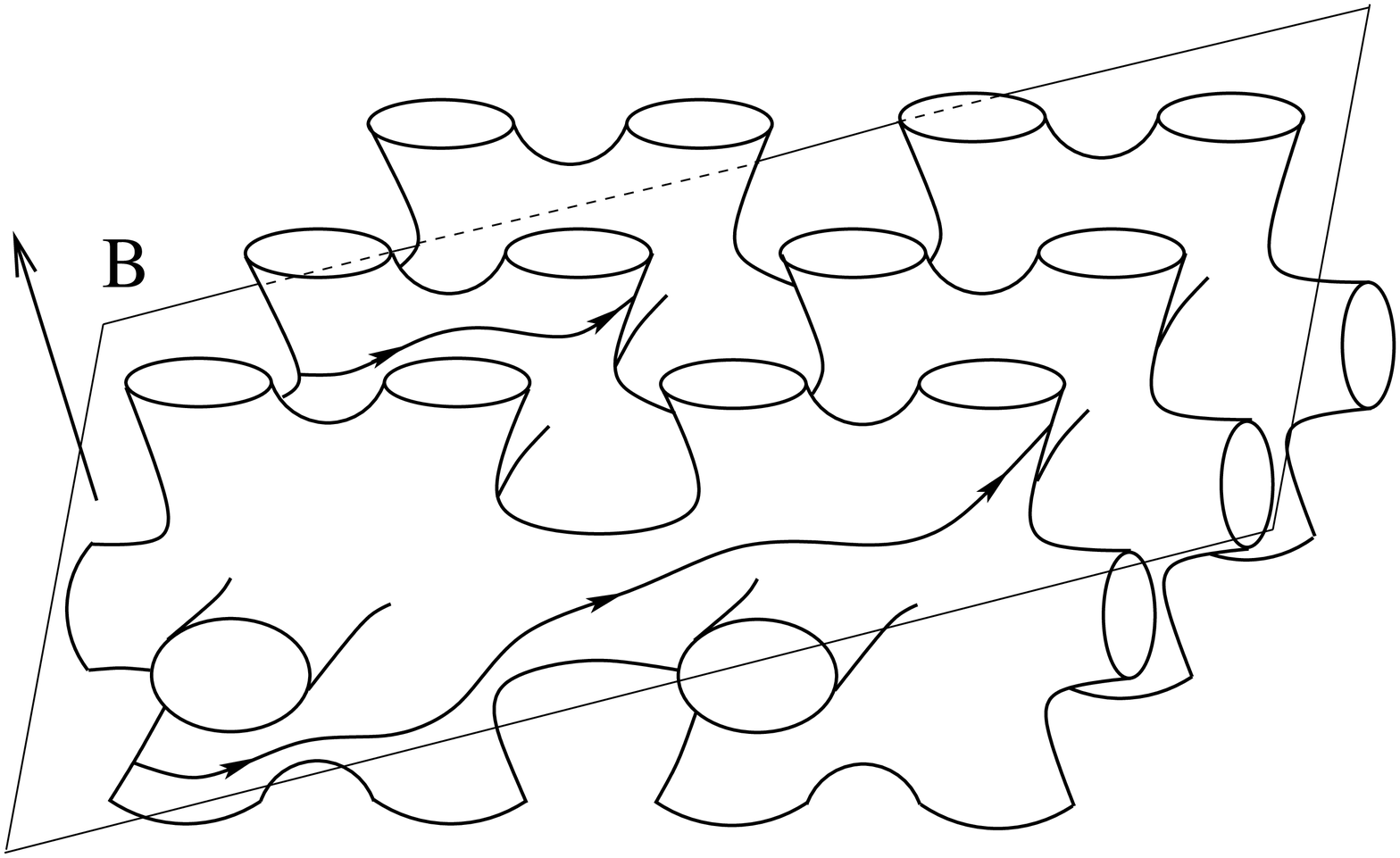}
\end{center}
\caption{Periodic surface in the three-space intersected by a plane 
of a given direction}
\label{PeriodSurf}
\end{figure}

 The most fundamental, from the point of view of physical applications, is 
Novikov's problem for functions with three quasi-periods. This problem was 
first set in \cite{MultValAnMorseTheory} and can also be considered as the problem 
of qualitative description of the geometry of intersections of an arbitrary 
two-dimensional periodic surface in $\mathbb{R}^{3}$ with a family of planes of 
a given direction (Fig.~\ref{PeriodSurf}).

 In this formulation, Novikov's problem is most directly related to the 
description of galvanomagnetic phenomena in metals in a constant uniform magnetic 
field at low temperature. The role of the periodic surface here is played by the 
Fermi surface in the space of quasimomenta. Intersections of the Fermi surface 
with planes orthogonal to the magnetic field determine the geometry of quasiclassical 
electron trajectories in this space, and the corresponding quasiperiodic functions 
$f (x, y)$ are given by the restrictions of a periodic (three-dimensional) 
dispersion relation $\epsilon (\mathbf p)$ to these planes. As was previously 
shown in a number of important examples (see~\cite{lifazkag,lifpes1,lifpes2,etm}), 
the behavior of transport phenomena (magnetic conductivity) in a metal in the limit 
of strong magnetic fields depends most significantly on the geometry of the described 
trajectories, which allows to conduct their experimental study. The most interesting 
magnetic transport phenomena are associated in this case with the presence of 
non-closed (open) electron trajectories on the Fermi surface; therefore, the most important is the 
classification of the open level lines of the corresponding functions $f (x, y)$.

To date, Novikov's problem has been studied most deeply in the case of three quasi-periods
(see \cite{zorich1,dynn1992,Tsarev,dynn1,DynnBuDA,dynn2,dynn3}). In particular, 
we have now a qualitative classification of open level lines of the functions $f (x, y)$,
which establishes their division into topologically regular \cite{zorich1,dynn1992,dynn1} and chaotic 
ones \cite{Tsarev,DynnBuDA,dynn2}. A detailed study of topologically regular level lines 
made it possible to introduce non-trivial topological characteristics associated with them, 
which were previously unknown. Such characteristics have the form of irreducible integer 
triples $(m^{1}, m^{2}, m^{3})$ and are defined for each stable family of topologically 
regular level lines (trajectories). In \cite{PismaZhETF,UFN}, these characteristics were 
defined as new topological numbers observed in the conductivity of normal metals with
complex enough Fermi surfaces.

 As for chaotic level lines of functions $f (x, y)$ with three quasi-periods, their 
existence was not known before~\cite{Tsarev,DynnBuDA,dynn2}. All such level lines are 
unstable with respect to small variations of the problem parameters, and their geometry 
is very complucated.

 Chaotic level lines can be divided into two main types, Tsarev-type trajectories 
and Dynnikov-type ones. The description of the global behavior of the trajectories 
in the second case is especially difficult, and the appearance of such trajectories on the 
Fermi surface leads to the most nontrivial, not considered earlier, behavior of the magnetic conductivity in 
strong magnetic fields (\cite{ZhETF2,TrMian}).

 The study of chaotic level lines of functions with three quasi-periods has been 
continued since their discovery (see, e.g.,
\cite{zorich2,Zorich1996,Zorich1997,ZorichAMS1997,zorich3,DeLeo1,DeLeo2,DeLeo3,
ZorichLesHouches,DeLeoDynnikov1,dynn4,DeLeoDynnikov2,Skripchenko1,Skripchenko2,DynnSkrip1,
DynnSkrip2,AvilaHubSkrip1,AvilaHubSkrip2}). In this article, we will try, in particular, 
to give the most detailed description of the current state of this area.

 The theory of galvanomagnetic phenomena in metals, however, is not the only field 
of physical applications of the general Novikov's problem. Naturally, numerous applications 
of the theory of quasi-periodic functions on the plane arise also in the physics of 
two-dimensional systems (see \cite
{Guidoni2,JMathPhys,SanchezPalenciaSantos,ViebSbrosCartYuSchneid,GautierYaoSanchezPalencia,
QuasiperGas}). As a rule, quasi-periodic functions $f (x, y)$ play the role of potentials 
in which the dynamics of particles localized in two dimensions is observed.
The applications of Novikov's problem here are related primarily with the description 
of transport phenomena in such systems (both in the presence of a magnetic field and without it). 
When describing such phenomena, the main role can be played by both the geometry of the level 
lines of the potential $f (x, y)$ (see e.g. \cite{JMathPhys}) and the geometry of the domains 
$f (x, y) \leqslant \epsilon_{0}$ (\cite{QuasiperGas}), which is directly related to the geometry 
of the level lines.

 In addition to purely applied significance, the study of Novikov's problem also has an important 
general theoretical one. Namely, quasi-periodic potentials with a 
sufficiently large number of quasi-periods on the plane can be considered as a transitional link 
between ``ordered'' and random potentials, having the properties of potentials of both types. 
To describe random potentials, various models are used, which may differ from each other 
in a number of properties. Some of the features of random potentials, however, are usually 
considered universal and related to the behavior of the potential level lines. In particular, 
random potentials are characterized by the presence of open level lines only at a single energy 
level $V (x, y) = \epsilon_{0}$, while at other energies all level lines are closed curves 
(see e.g. \cite{Stauffer,Essam,Riedel,Trugman}). Open level lines of random potentials, 
as a rule, have a rather complex geometry, wandering around the plane in a chaotic manner.
 
 Considering Novikov's problem from the point of view of random potential models, already 
in the case of three quasi-periods, one can observe both rich families of ``regular'' potentials 
(having topologically regular open level lines in a finite energy interval) and non-trivial 
examples of ``random'' potentials (having chaotic level lines present only at a single energy level).

 Experimental techniques, as a rule, make it possible to create families of quasi-periodic 
potentials with a given number of quasi-periods, depending on a finite number of parameters 
$\mathbf U = (U^{1}, \dots, U^{N})$. In most of these cases, the results of the study of
Novikov's problem for three quasi-periods lead to a universal (albeit rather nontrivial) 
description of the sets of ``regular'' and ``random'' potentials within the full family 
$V (x, y, \mathbf U)$.

 Namely, the parameter space contains an everywhere dense set consisting of domains with 
piecewise smooth boundaries, each of which is a ``stability zone'' and corresponds to 
potentials with topologically regular level lines. Each of the stability zones is determined 
by its own values of the topological invariants $(m^{1}, m^{2}, m^{3})$. The complement to 
the union of all stability zones in the parameter space $\mathbf U$ is a set of fractal type 
and parametrizes potentials with chaotic level lines. It is this set that can be considered 
in this case as a realization of the model of quasi-periodic potentials with the properties 
of random potentials.
 
 As the study of Novikov's problem with four quasi-periods \cite{NovKvazFunc,DynNov} shows, 
a natural division of the set of potentials into subsets of potentials with topologically 
regular and chaotic open level lines also arises here. As in the case of three quasi-periods, 
topological invariants are associated with topologically regular open level lines, and have 
now the form of irreducible integer quadruples $(m^{1}, m^{2}, m^{3}, m^{4})$.

 As follows from the results of \cite{NovKvazFunc,DynNov}, smooth families of quasi-periodic 
potentials $V (x, y, \mathbf U)$ with four quasi-periods must in the generic case also contain 
an everywhere dense set, which is the union of stability zones, corresponding to potentials with 
topologically regular level lines, as well as a fractal complement to this set that parametrizes 
potentials with chaotic level lines. The question of whether chaotic level lines are present 
in a certain nondegenerate energy interval or only at a single energy level remains open for 
potentials with four quasi-periods. Note that it is the potentials with four quasiperiods 
that are most closely related to the theory of two-dimensional quasicrystals, which we have 
already mentioned above.
 
  As for functions with a larger number of quasi-periods, at the moment there are practically 
no rigorous general results for them. For any number of quasi-periods, it is easy to construct 
functions that have stable topologically regular level lines. However, the question of whether 
they will be everywhere dense in smooth families of quasi-periodic potentials 
$V (x, y, \mathbf U)$ for~$N>4$ is still open. Also, at the moment there are no rigorous 
results on the description of chaotic level lines of such potentials. In this article, we 
describe the situation that arises here with the help of a number of examples, and also 
formulate and prove a number of general statements about level lines of functions with 
an arbitrary number of quasi-periods.

\section{Novikov's problem in the case of three quasi-periods and angular diagrams 
of magnetic conductivity in metals}

 In this section, we will stop in detail on Novikov's problem with three quasi-periods 
and its main application, the description of galvanomagnetic phenomena in metals in the 
presence of strong magnetic fields. Many key consequences of the results of the study of 
Novikov's problem for the theory of galvanomagnetic phenomena were revealed and presented 
in a number of papers already some time ago (see, for example,
\cite{PismaZhETF,UFN,ZhETF2,DynSyst,BullBrazMathSoc,JournStatPhys}).
After that, however, a number of new important aspects were revealed, which essentially 
supplemented the general picture both in terms of rigorous mathematical results and in 
the field of applications.

 In the setting described, the role of the periodic function in the ambient space is 
played by the dispersion relation $\epsilon (\mathbf p)$ defined in the space of 
quasi-momenta $\mathbf p = (p_{1}, p_{2}, p_{3})$. The function $\epsilon (\mathbf p)$ 
is periodic with respect to the reciprocal lattice, whose basis vectors 
$\mathbf a_{1} $, $\mathbf a_{2} $, $\mathbf a_{3}$ are connected with the basis of the 
crystal lattice $( \mathbf l_{1},\mathbf l_{2}, \mathbf l_{3} )$ by the relations
$$\mathbf a_{1} = 2 \pi \hbar
{\mathbf l_{2} \times \mathbf l_{3} \over (\mathbf l_{1}, \mathbf l_{2}, \mathbf l_{3})} 
,  \quad
\mathbf a_{2} = 2 \pi \hbar
{\mathbf l_{3} \times \mathbf l_{1} \over (\mathbf l_{1}, \mathbf l_{2}, \mathbf l_{3})} 
,  \quad
\mathbf a_{3} = 2 \pi \hbar
{\mathbf l_{1} \times \mathbf l_{2} \over (\mathbf l_{1}, \mathbf l_{2}, \mathbf l_{3})} $$

 In the presence of an external magnetic field, a nontrivial semiclassical dynamics of 
electronic states arises in the space of quasimomenta, which is determined by the system
\begin{equation}
\label{MFSyst}
\dot{\mathbf p} = {e \over c}
\left[ \mathbf v_{\mathrm{gr}} \times \mathbf B \right] =
{e \over c}\left[ \nabla \epsilon (\mathbf p) \times \mathbf B \right] 
, 
\end{equation}
(see e.g. \cite{etm,Kittel,Ziman,Abrikosov}).

 Geometrically, the trajectories of the system \eqref{MFSyst} are given by the intersections 
of surfaces of constant energy $\epsilon (\mathbf p) = {\rm const}$ by planes orthogonal to 
the magnetic field, or, in other words, by the level lines of the function 
$\epsilon (\mathbf p) $ restricted to such planes. For a given dispersion relation 
$\epsilon (\mathbf p)$, we thus have a family of quasi-periodic functions on the plane, 
the parameters in which are given by the direction of the magnetic field $\mathbf B$ and 
the shift of the plane relative to the origin. (Moreover, for the questions of interest 
to us, the shift does not play a role in the case of generic direction $\mathbf B$.)

 Many of our results are formulated for \emph{generic} functions. This means that 
the function belongs to some fixed open everywhere dense subset in the space of all smooth 
functions.

 The trajectories of system \eqref{MFSyst} can, of course, be both closed and open 
in the $\mathbf p$-space. The following property of closed trajectories is specific for 
the level lines of quasi-periodic functions with three quasi-periods.

\begin{lemm}[\cite{dynn1}]\label{lem1}
For any fixed direction $\mathbf B$ and energy value $\epsilon(\mathbf p)=\epsilon_0$, 
the diameters of all closed trajectories of the system~\eqref{MFSyst} in the $\mathbf p$-space 
are bounded by one constant (depending on $\mathbf B$ and $\epsilon_0$).
\end{lemm}

 The value of the corresponding constant, however, may depend on the direction of 
$\mathbf B$ and on $\epsilon_{0}$, becoming arbitrarily large as these parameters change. 
As we have already said, we will be interested in the phenomena associated with the 
presence of non-closed trajectories of the system \eqref{MFSyst}.

 The contribution to the magnetic conductivity comes from trajectories in all planes 
orthogonal to the magnetic field. However, if the direction of $\mathbf B$ is not 
proportional to an integral one (that is, to a direct lattice vector), then the 
image of any plane orthogonal to $\mathbf B$ is everywhere dense in the 
torus~$\mathbb T^3=\mathbb R^3/\mathbb Z^3$. Therefore, in different planes orthogonal 
to $\mathbf B$ the level lines of the dispersion relation~$\epsilon$ behave similarly.

\begin{figure}[t]
\begin{center}
\includegraphics[width=0.9\linewidth]{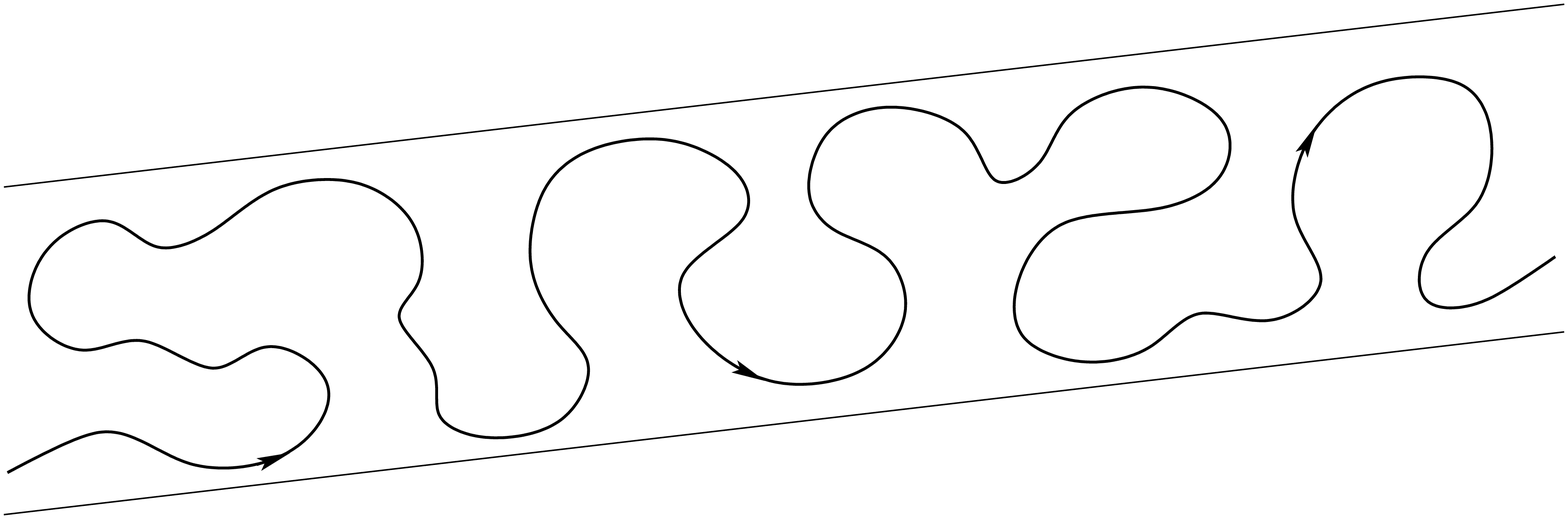}
\end{center}
\caption{Topologically regular open trajectory of the system \eqref{MFSyst} 
in the plane orthogonal to $\mathbf B$ (schematically)}
\label{StableTr}
\end{figure}

 We divide open trajectories into two types, \emph{topologically regular} and \emph{chaotic}. 
Topologically regular are open trajectories whose projection to~$\mathbb T^3$ is contained 
in an embedded two-dimensional torus. The \emph{topological characteristics} of such 
a trajectory include the possible homology classes that this torus may have. 
(If the trajectory is not periodic, then such a homology class is uniquely defined 
up to a factor.) Any topologically regular trajectory lies in a straight strip of 
finite width in some plane orthogonal to $\mathbf B$, and passes through it 
(Fig.~\ref{StableTr}). The direction of the strip is determined by the 
topological characteristics of the trajectory.

 An open trajectory that is not topologically regular is called chaotic.

\begin{lemm}[\cite{dynn2}]
For any fixed direction $\mathbf B$ not proportional to an integral one, all open 
trajectories of the system \eqref{MFSyst} in all planes orthogonal to $\mathbf B$ have 
the same type and topological characteristics, which also do not depend on energy 
values $\epsilon_{0}$ \emph(provided that not all trajectories are closed at this level\emph).
\end{lemm}

 If the direction of $\mathbf B$ is integral, then the open trajectories of the 
system \eqref{MFSyst} have a relatively simple description, namely, they can only be 
periodic.

 We will divide directions $\mathbf B$ into rational (proportional to integral ones), 
partially irrational (such that the plane orthogonal to $\mathbf B$ contains only one reciprocal 
lattice vector up to a factor), and generic directions (the plane orthogonal to 
$\mathbf B$ does not contain nonzero reciprocal lattice vectors).

\begin{figure}[t]
\begin{tabular}{cc}
\includegraphics[width=0.45\linewidth]{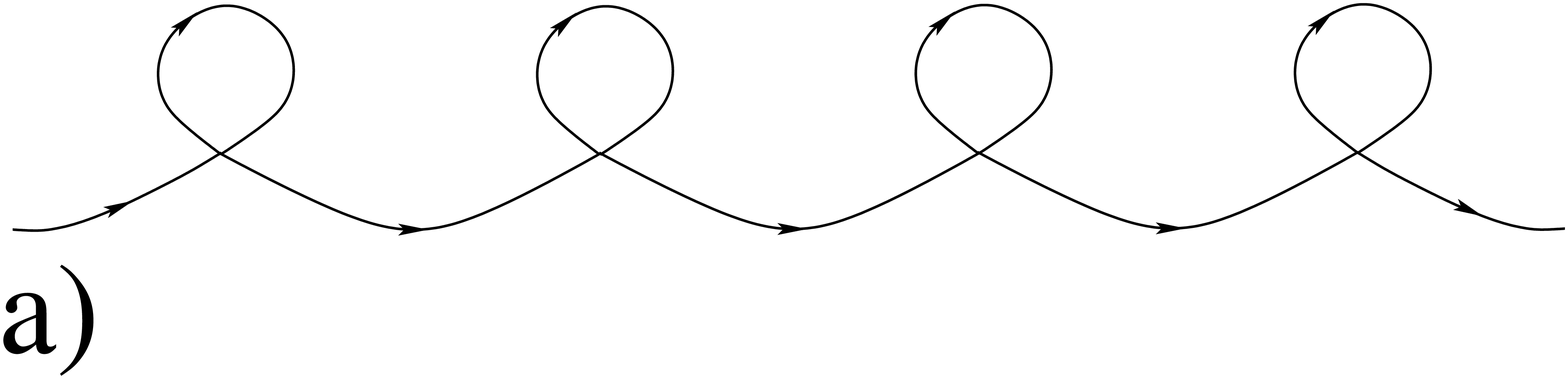}  &
\hspace{5mm}
\includegraphics[width=0.45\linewidth]{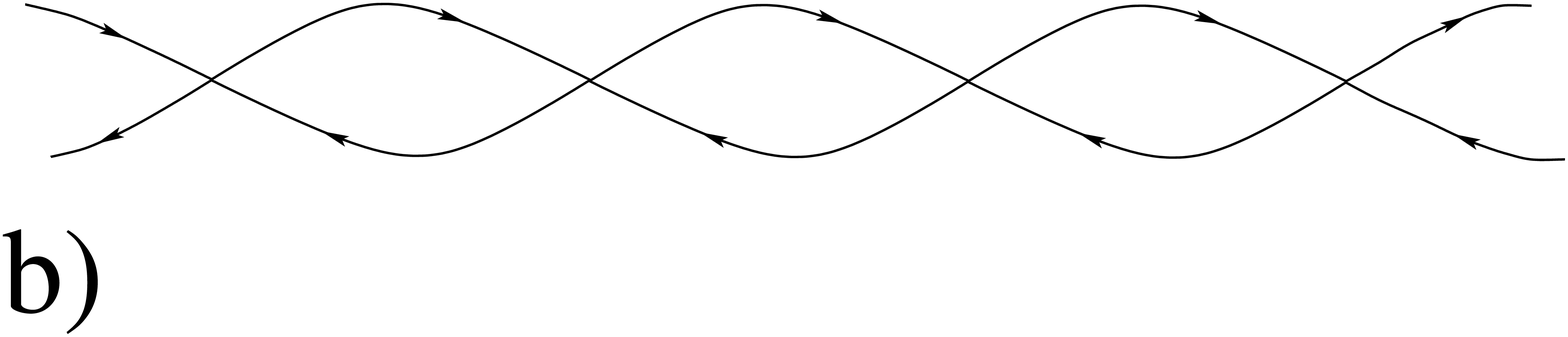}
\end{tabular}

\vspace{5mm}

\begin{tabular}{cc}
\includegraphics[width=0.15\linewidth]{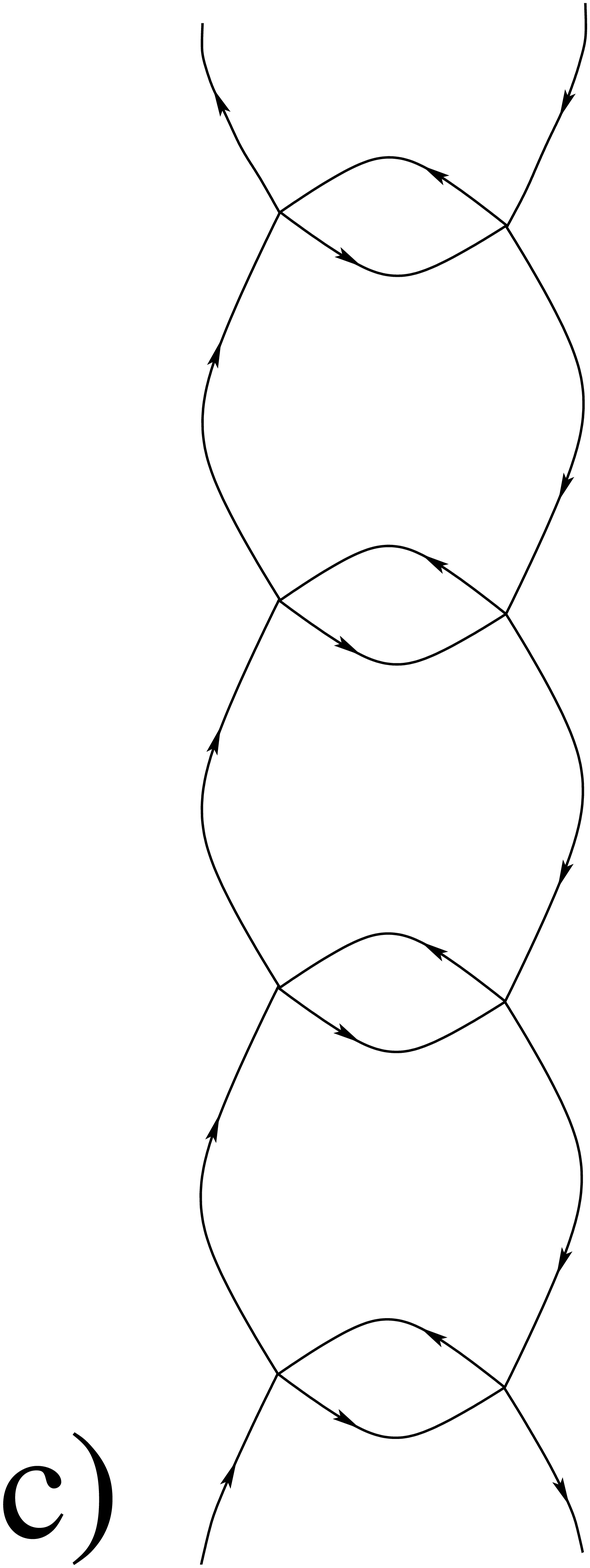}
\hspace{10mm}
\includegraphics[width=0.75\linewidth]{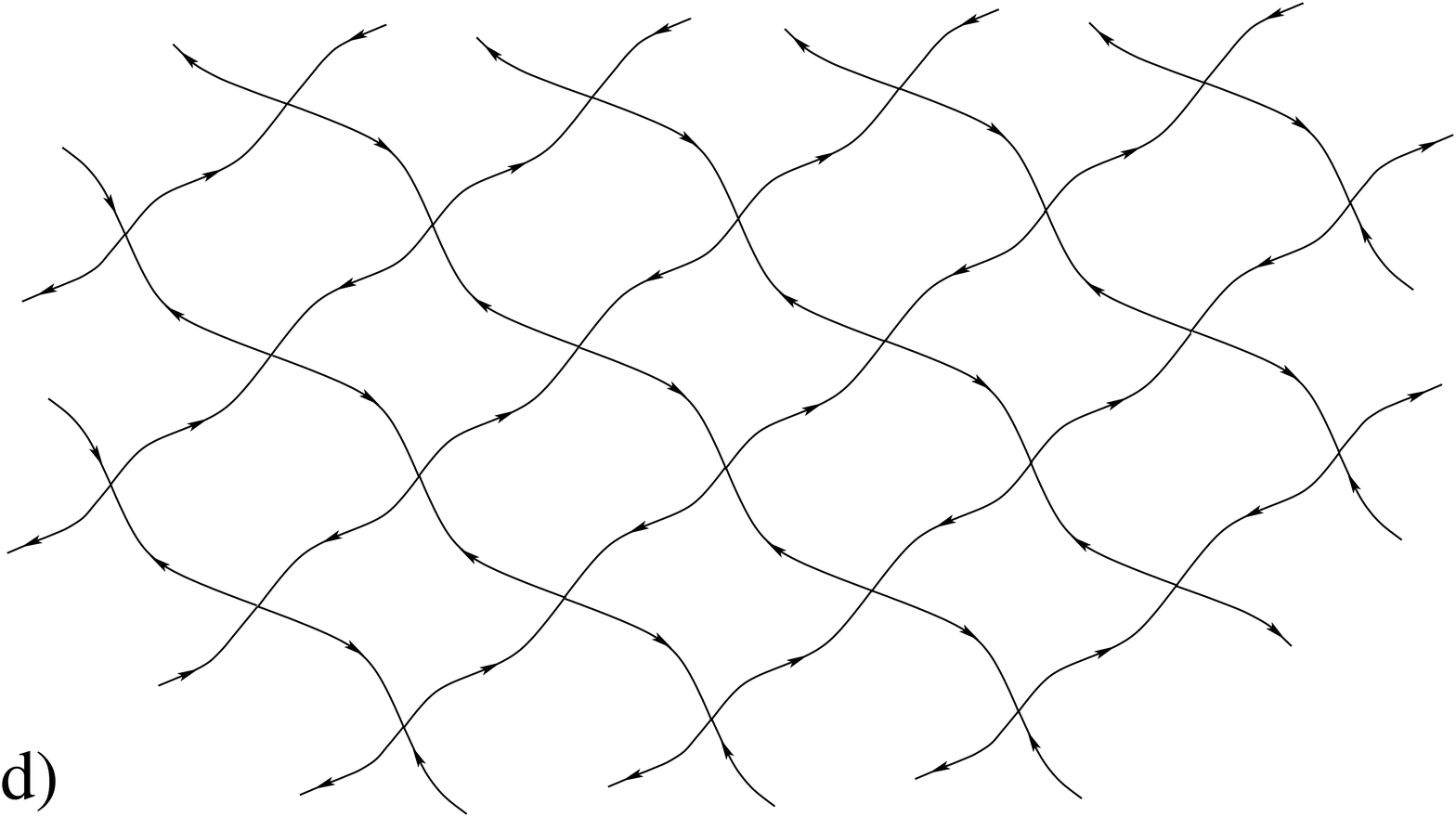}
\end{tabular}
\caption{Connected complexes of stationary points and separatrices of the 
system \eqref{MFSyst} unbounded in the $\mathbf p$-space}
\label{PeriodicSingular}
\end{figure}

\begin{figure}[t]
\begin{center}
\includegraphics[width=0.9\linewidth]{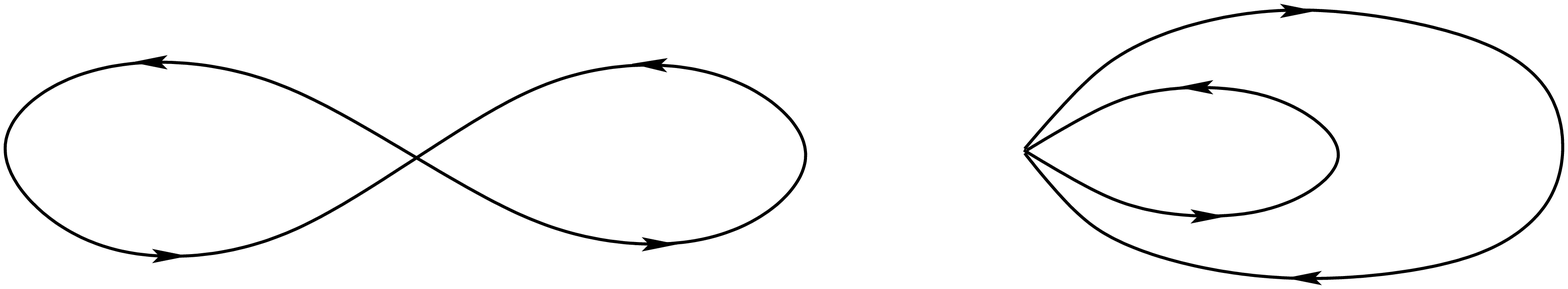}
\end{center}
\vspace{3mm}
\begin{tabular}{cc}
\hspace{5mm}
\includegraphics[width=0.45\linewidth]{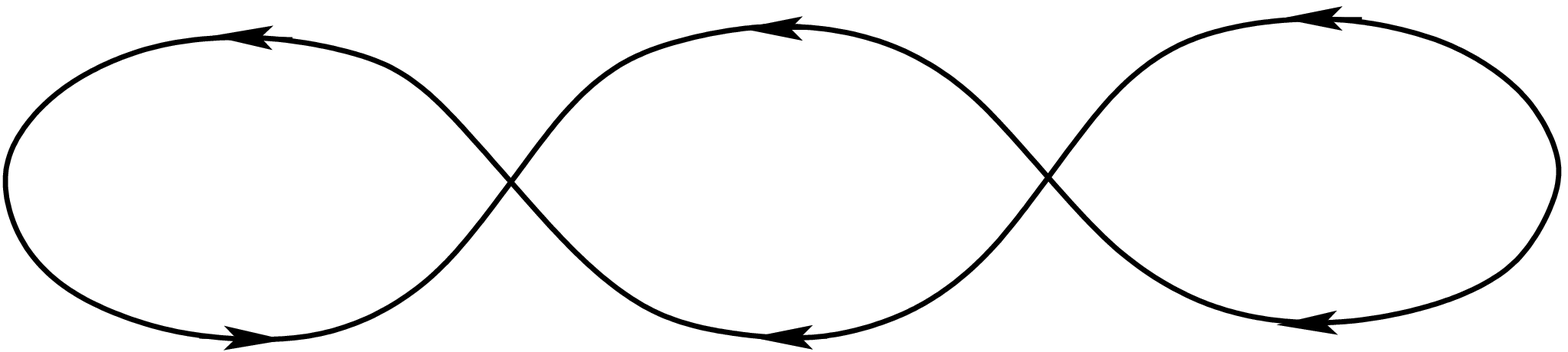}  &
\hspace{5mm}
\includegraphics[width=0.35\linewidth]{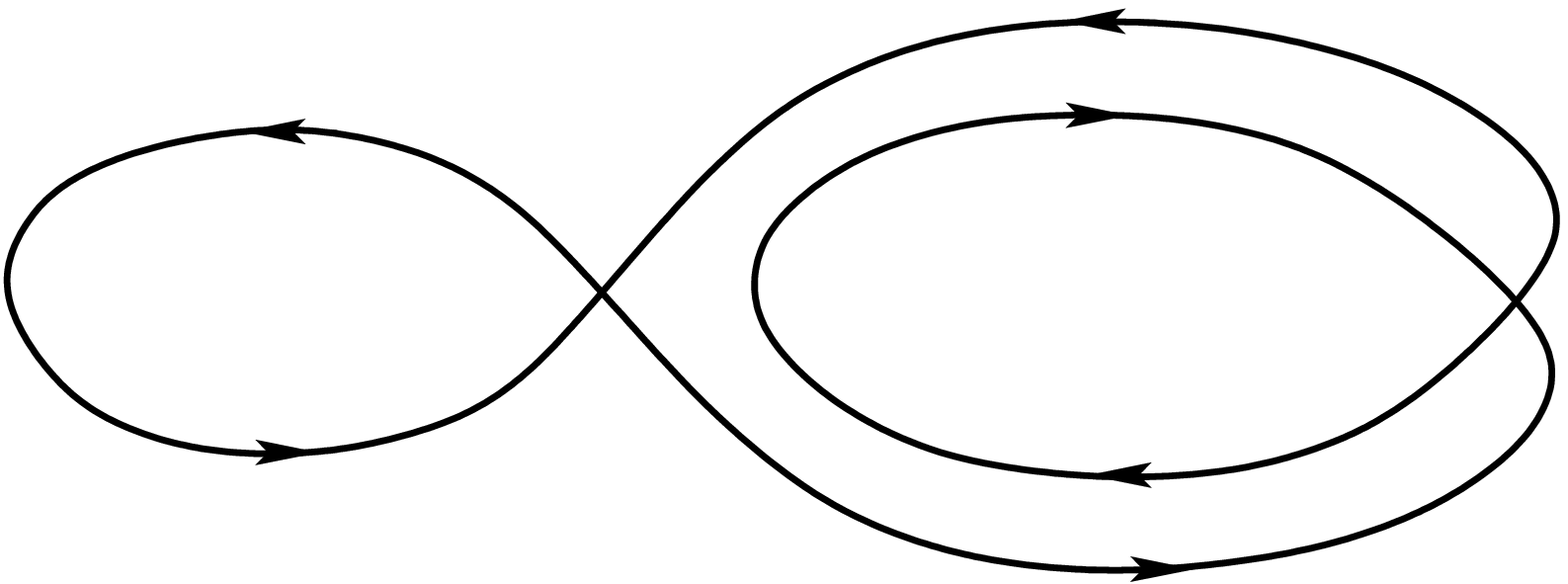}
\end{tabular}
\vspace{5mm}
\begin{tabular}{cc}
\hspace{5mm}
\includegraphics[width=0.35\linewidth]{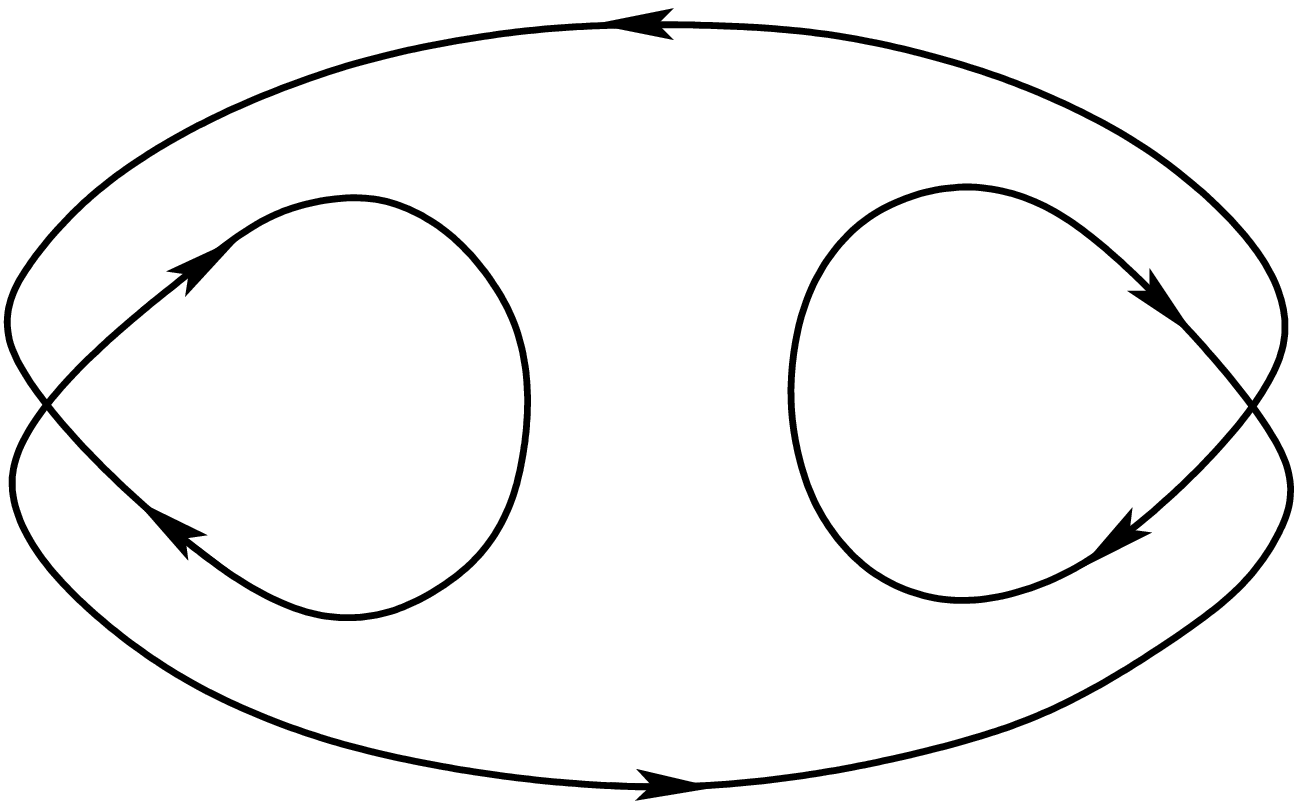}  &
\hspace{5mm}
\includegraphics[width=0.45\linewidth]{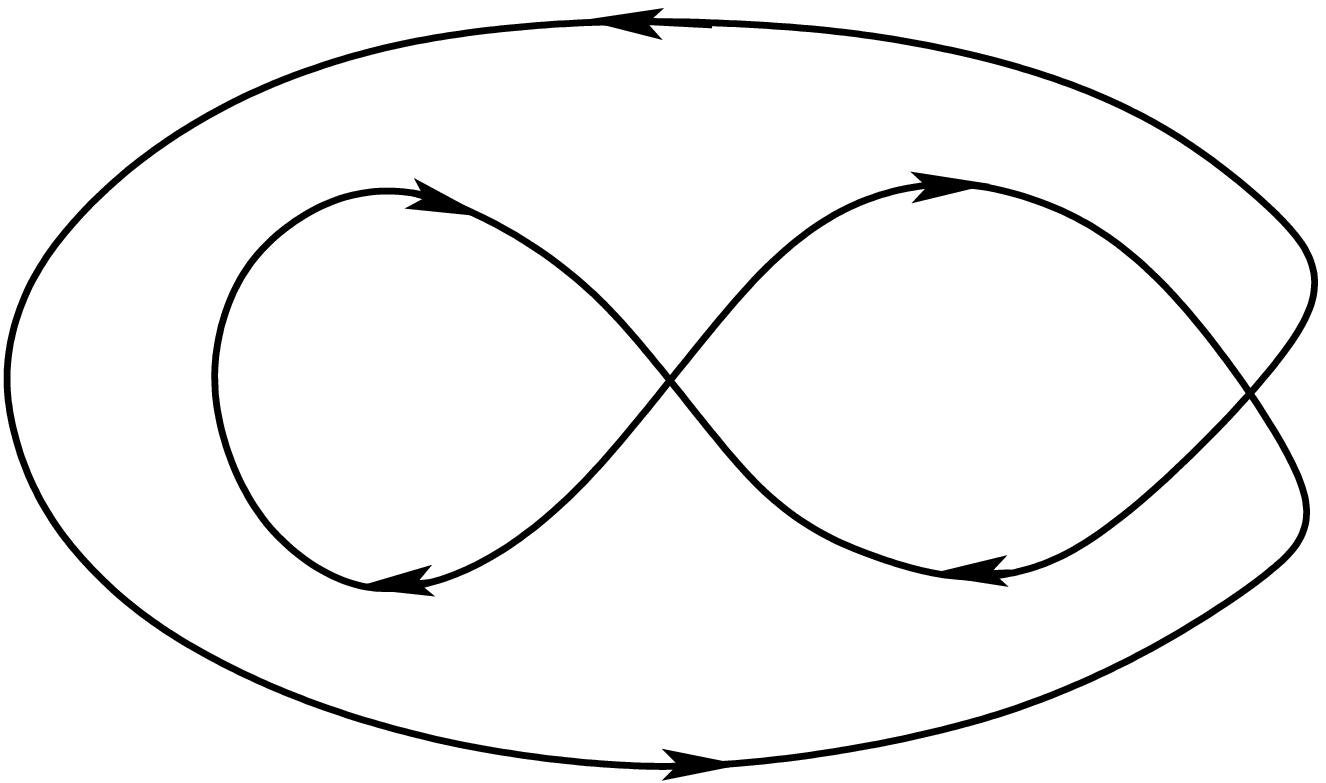}
\end{tabular}
\begin{tabular}{cc}
\hspace{5mm}
\includegraphics[width=0.45\linewidth]{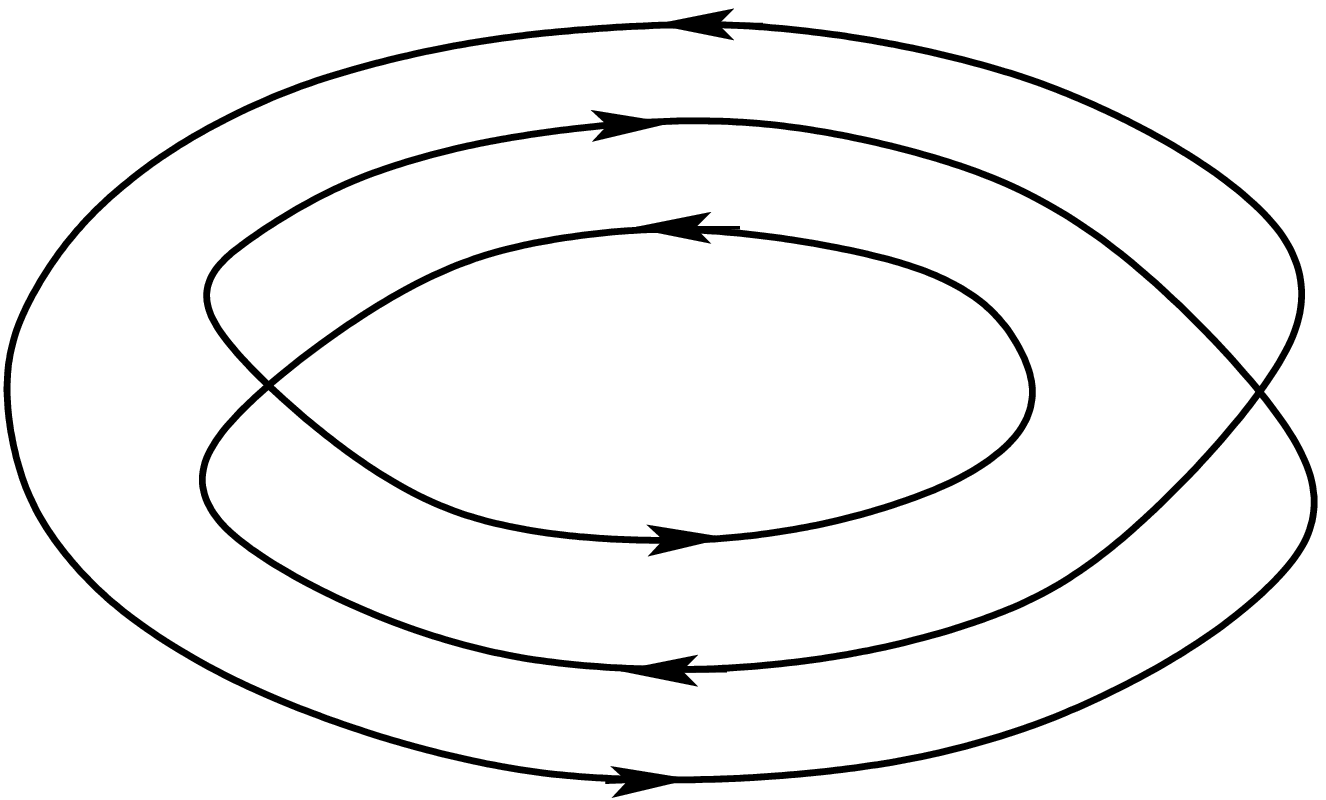}  &
\hspace{5mm}
\includegraphics[width=0.35\linewidth]{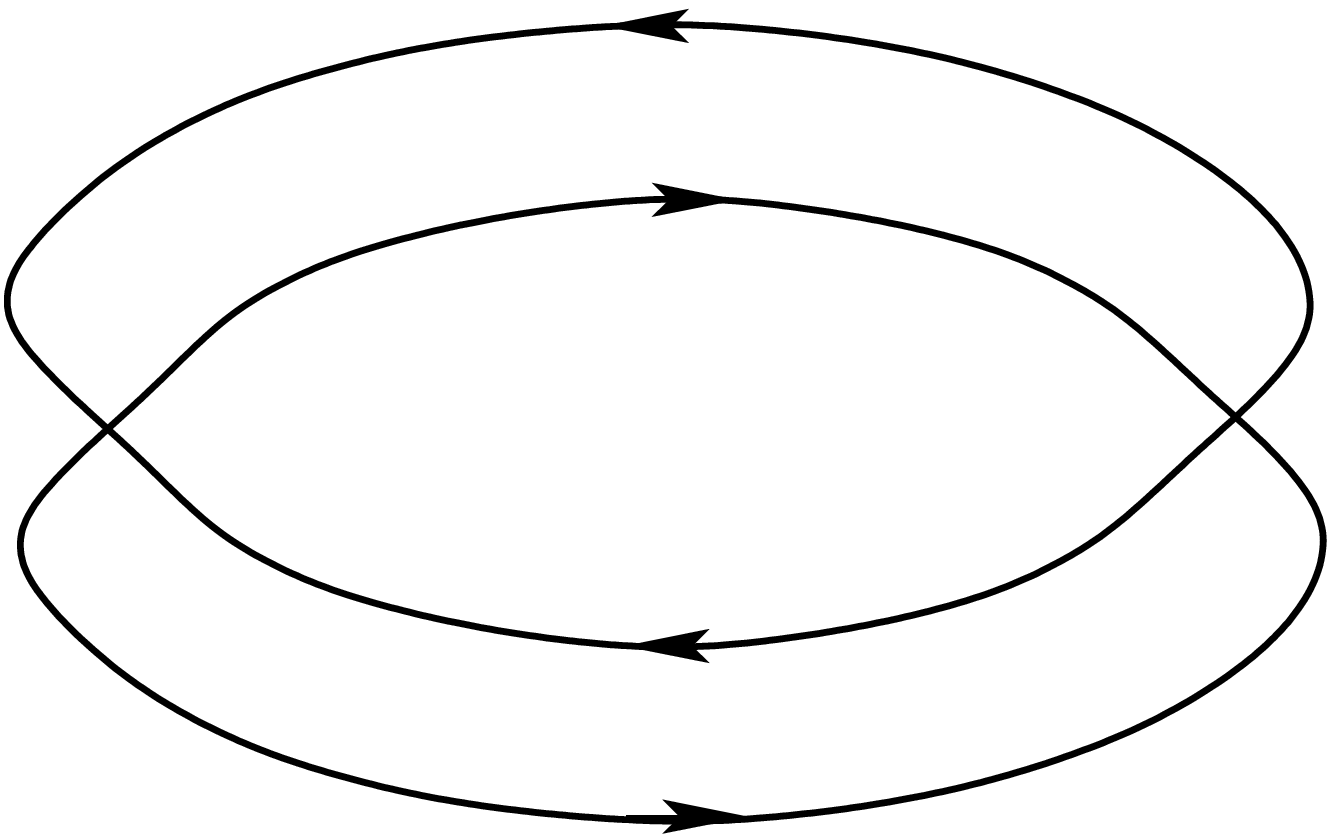}
\end{tabular}
\caption{Examples of connected complexes of stationary points and sepatarices 
of system \eqref{MFSyst} bounded in the $\mathbf p$-space}
\label{BoundedSingular}
\end{figure}

 In our situation it is natural to introduce an angular diagram showing the dependence of 
the type of open trajectories of system \eqref{MFSyst} on the direction of ${\bf B}$.  
For simplicity, here we will call open trajectories not only non-closed non-singular 
trajectories of the system \eqref{MFSyst}, but also connected complexes consisting of 
stationary points and separatrices connecting them provided they are not bounded in 
the $\mathbf p$-space (Fig. ~\ref{PeriodicSingular}). Similarly, in addition to 
closed non-singular trajectories of the system \eqref{MFSyst}, we will also call closed 
trajectories bounded in the $\mathbf p$-space connected complexes consisting of stationary points and separatrices connecting 
them (Fig.~\ref{BoundedSingular}). With this definition, 
Lemma~\ref{lem1} is still valid, and the following assertion also holds.

\begin{lemm}[\cite{dynn2,dynn3}]\label{lem3}
 For any fixed direction $\mathbf B$, the set of energy values~$\epsilon$ for 
which the system \eqref{MFSyst} has open trajectories has the form of a segment
$[\epsilon_{1} (\mathbf B) ,
\epsilon_{2} (\mathbf B) ]$,
which can degenerate into a single point
$\epsilon_{0} (\mathbf B) = \epsilon_{1} (\mathbf B) 
= \epsilon_{2} (\mathbf B)$. 
\end{lemm}

 The main feature of the angular diagram describing the open trajectories of 
system \eqref{MFSyst} is the presence of an everywhere dense set having the form of a union 
of ``stability zones'' each of which is a domain with a piecewise-smooth boundary on the 
unit sphere and includes directions $ \mathbf B$ corresponding to topologically regular 
open trajectories (see~\cite{zorich1,dynn1992,dynn1,dynn2,dynn3}). Namely, the following 
assertion is true.

\begin{theo}\label{everywheredense-th}
For every generic dispersion law $\epsilon$, there is an everywhere dense set on the 
sphere~$\mathbb S^2$, which is the union of at most countably many closed 
domains~$\Omega_\alpha$ with the following properties\emph:
\begin{enumerate}
\item the interiors of the domains~$\Omega_\alpha$ are pairwise disjoint\emph;
\item if~$\mathbf B\in\bigcup_\alpha\Omega_\alpha$, then the trajectories of the 
system~\eqref{MFSyst} are topologically regular\emph;
\item
each stability zone~$\Omega_\alpha$ corresponds to an irreducible triple of integers 
$(m^{1}_{\alpha}, m^{2}_{\alpha}, m^{3}_{\alpha}) $, characterized uniquely up to sign 
by the property that for~$\mathbf B\in\Omega_\alpha$ any open trajectory of 
the system~\eqref{MFSyst} lies in a strip whose direction is orthogonal to the vector 
with coordinates~$(m^{ 1}_{\alpha}, m^{2}_{\alpha}, m^{3}_{\alpha})$ with respect to 
the basis of the crystal lattice\emph;
\item if~$\mathbf B\notin\bigcup_\alpha\Omega_\alpha$ and the direction of $\mathbf B$ 
is completely irrational, then open trajectories of the system~\eqref{MFSyst} are 
present only at one energy level, i.e.~$\epsilon_1(B)=\epsilon_2(B)$, and are chaotic.
\end{enumerate}
\end{theo}

 Thus, for a fixed generic direction $\mathbf B\in\Omega_\alpha$, the topologically 
regular trajectories of system \eqref{MFSyst} have the same mean direction given by the 
intersection of the plane orthogonal to $\mathbf B$ with some integral 
plane $\Gamma_{\alpha}$, which is the same for a given stability zone $\Omega_{\alpha}$. 
This plane is orthogonal to the vector
$$\mathbf l_{\alpha} = m^{1}_{\alpha}\mathbf l_{1}
+m^{2}_{\alpha}\mathbf l_{2}+
m^{3}_{\alpha}\mathbf l_{3} $$
of the crystal lattice. 

 In the general case, topologically regular open trajectories of \eqref{MFSyst} are 
in some sense quasi-periodic. However, each stability zone~$\Omega_\alpha$ contains 
an infinite set of directions $\mathbf B$ for which the open trajectories of the 
system~\eqref{MFSyst} are periodic. This happens whenever the intersection of the 
plane orthogonal to $\mathbf B$ and~ $\Gamma_{\alpha}$ has a rational direction 
in the $\mathbf p$-space.

 Note that topologically regular trajectories are naturally divided into stable, 
unstable, and semi-stable ones. Namely, a topologically regular trajectory is 
called \emph{stable} if for any of its points and any open neighborhood~$V$ of 
this point, for any sufficiently small perturbation of the dispersion law~$\epsilon$, 
energy level~$\epsilon_0$, and direction of magnetic field~$\mathbf B$, some topologically 
regular trajectory of the perturbed system intersects~$V$. If, by an arbitrarily small 
perturbation of the system, we can achieve that a stable topologically regular trajectory 
passes through~$V$, then the original trajectory is said to be \emph{semi-stable}. 
In other cases, topologically regular trajectories are called \emph{unstable}.

 The remarkable properties of topologically regular open trajectories of system 
\eqref{MFSyst} made it possible to introduce in \cite{PismaZhETF,UFN} new topological 
characteristics, observable in the conductivity of normal metals whenever stable 
trajectories of this type are present on the Fermi surface. The possibility of 
introducing such characteristics is based on the features of the contribution of 
such trajectories to the conductivity tensor in the limit of strong magnetic fields, 
among which the most important is the strong anisotropy of the conductivity in the 
plane orthogonal to~$\mathbf B$. Measuring the conductivity in this plane allows to 
directly measure the mean direction of the open trajectories in 
the $\mathbf p$-space as the direction of the greatest suppression of the conductivity 
in the limit $B \rightarrow \infty$. In the presence of open topologically regular 
trajectories on the Fermi surface that are stable under small rotations of
$\mathbf B$, the measurement of conductivity in strong magnetic fields thus allows 
to determine the integer invariants 
$(m^{1}_{\alpha}, m^{2}_{\alpha}, m^{3}_{\alpha})$ corresponding to the given family 
of open trajectories.

 As follows from~\cite{dynn3}, the complete angular diagrams for periodic dispersion 
relations $\epsilon (\mathbf p)$ can belong to only one of the following two types.
\begin{theo}
For a generic dispersion law~$\epsilon$ one of the following two cases occurs\emph:
\begin{enumerate}
\item
the entire sphere~$\mathbb S^2$ is the only stability zone\emph;
\item
the angular diagram for~$\epsilon$ contains an infinite number of 
stability zones~$\Omega_\alpha$, with numbers 
$(m^{1}_{\alpha}, m^{2}_{\alpha}, m^{3}_{\alpha})$ growing without bound in absolute value.
\end{enumerate}
\end{theo}

\begin{figure}[t]
\begin{center}
\includegraphics[width=0.9\linewidth]{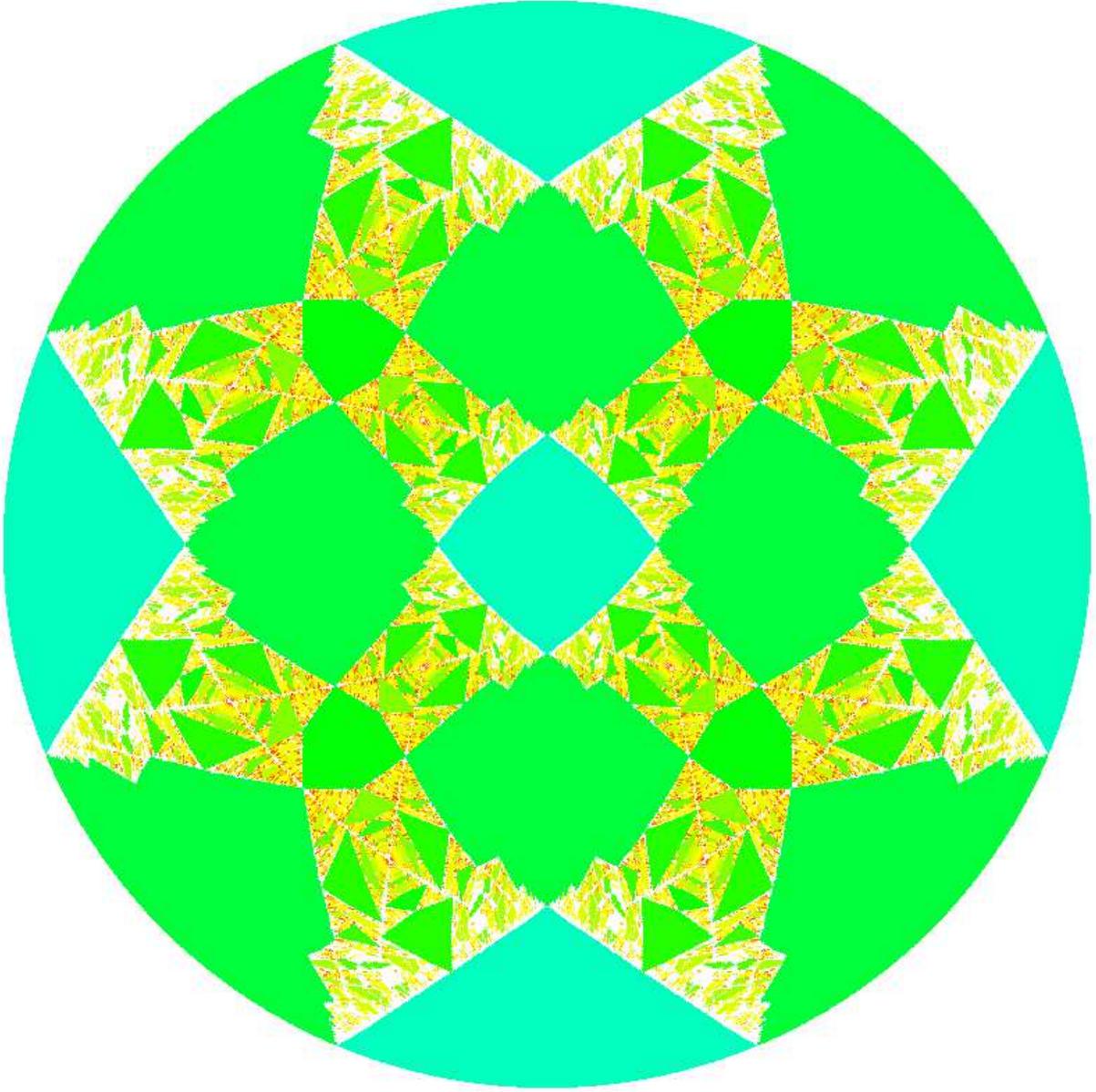}
\end{center}
\caption{Stability zones on a sphere for the dispersion relation
$\epsilon (\mathbf p)=\cos p_{x}\cos p_{y}+
\cos p_{y}\cos p_{z}+\cos p_{z}\cos p_{x}$ 
(in the stereographic projection)
\cite{DeLeoObzor}
\label{cos3DmDisc}}
\end{figure}

 Fig.~\ref{cos3DmDisc} shows an example of an angle diagram of the second type. 
The next statement, from which the previous theorem follows, describes a key property 
of such diagrams, namely, their structure near the boundary of each stability zone.

\begin{figure}[t]
\includegraphics[width=\linewidth]{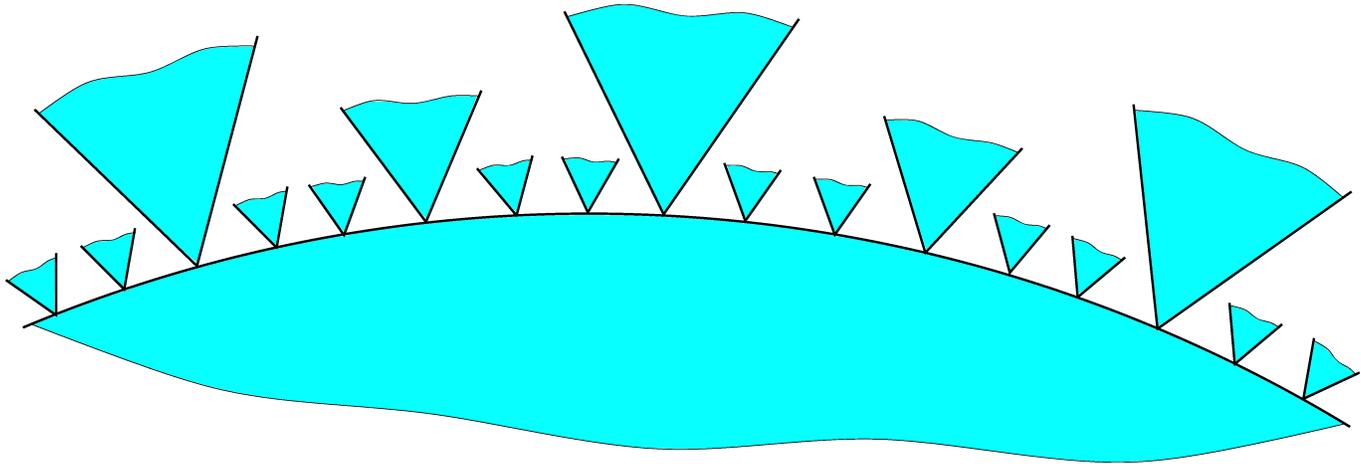} 
\caption{An infinite set of stability zones $\Omega_{\beta}$ adjacent to the boundary
of a zone $\Omega_{\alpha}$}
\label{Adjacent}
\end{figure}

\begin{theo}
Let~$\mathbf B$ be a point of smoothness of the boundary of some stability zone 
$\Omega_{\alpha}$ such that all open trajectories of the system \eqref{MFSyst} 
are periodic. Then~$\mathbf B$ also belongs to the boundary of another zone, 
$\Omega_{\beta}$, such that~$\partial\Omega_\beta$ has a breaking point at~$\mathbf B$, 
and~$\mathbf B$ is an isolated intersection point of~$ \Omega_\alpha$ and~$\Omega_\beta$. 
The set of such points is 
dense in the boundary of each stability zone \emph(see~Fig.~\ref{Adjacent}\emph).
\end{theo}

One can see that the presence of periodic trajectories of the 
system~\eqref{MFSyst} means that the direction of $\mathbf B$ is orthogonal to 
some vector of the reciprocal lattice, to the same one under shift by which 
the trajectories are invariant. This means that~$\mathbf B$ is contained in 
some plane generated by two vectors of the direct lattice. If~$\mathbf B\in\Omega_\alpha$, 
then the vector~$\mathbf l_\alpha$ corresponding 
to the given stability zone  can be taken for one of these vectors, since the mean direction 
of the open trajectories is orthogonal to it.

\begin{figure}[t]
\includegraphics[width=\linewidth]{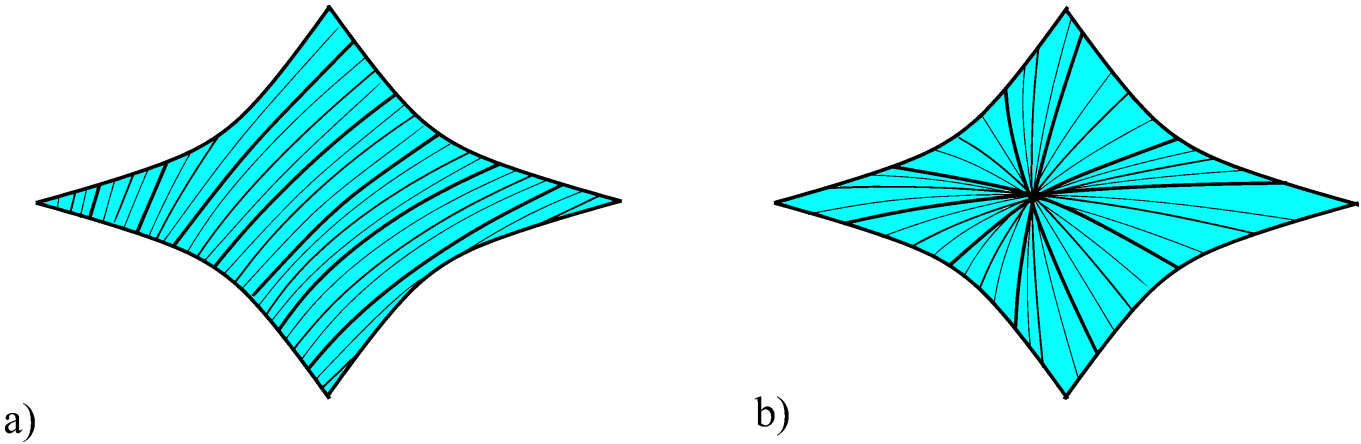} 
\caption{Everywhere dense families of directions $\mathbf B$ in stability zones for which 
the system~\eqref{MFSyst} has periodic open trajectories, in the cases where the zone does 
not contain (a) and contains (b) the vector~$\mathbf l_\alpha $}
\label{DenseNet}
\end{figure}

\begin{figure}[t]
\includegraphics[width=0.9\linewidth]{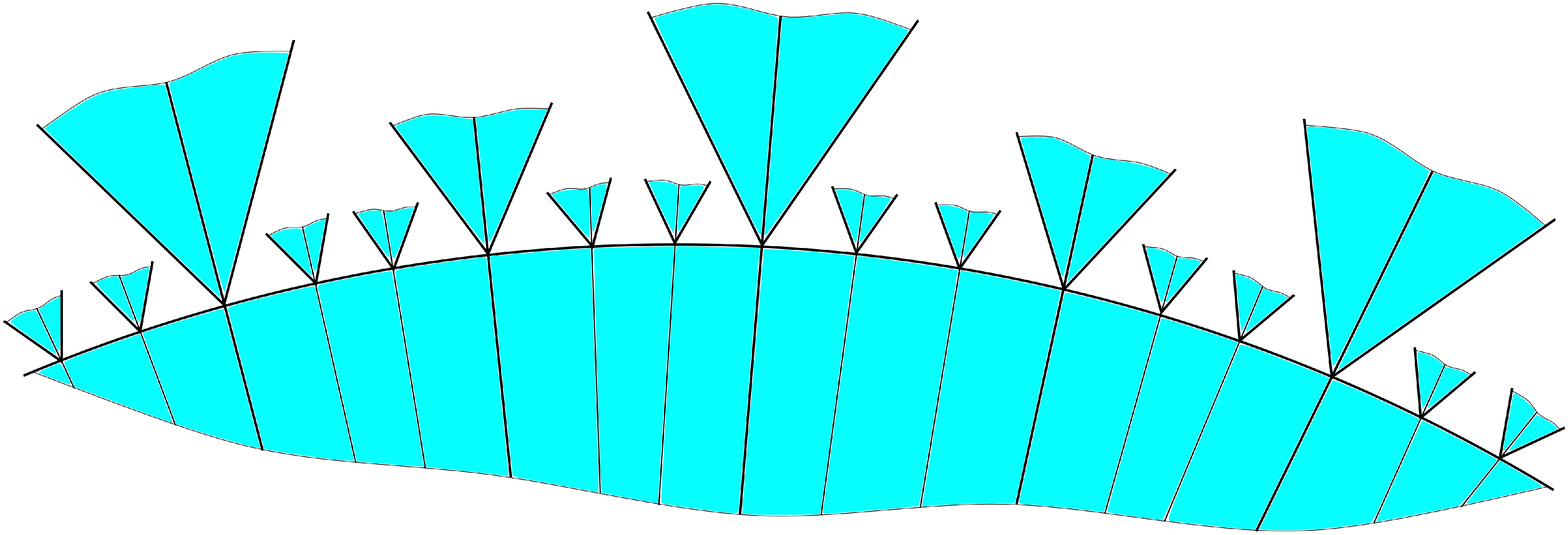} 
\caption{Curves of partially irrational directions $\mathbf B$ corresponding to 
the presence of periodic trajectories of \eqref{MFSyst} in adjacent stability zones}
\label{AdjacentLines}
\end{figure}

 Thus, the directions~$\mathbf B\in\Omega_\alpha$ for which the system~\eqref{MFSyst} 
has periodic trajectories form, on the 
angle diagram, an everywhere dense union of arcs (geodesics) passing through 
the point~$\mathbf l_\alpha$ (see Fig.~\ref{DenseNet}). 
The points at which these arcs intersect the boundary of the zone $\partial\Omega_\alpha$, 
excluding the breaking points of the boundary, are ones at which other 
stability zones are attached. Moreover, if~$\mathbf B\in\Omega_\alpha\cap\Omega_\beta$, 
then the vector~$\mathbf B$ is a linear combination of the vectors~$\mathbf l_\alpha$ 
and~$\mathbf l_\beta$, and the corresponding arc of the geodesic passing 
through~$\mathbf l_\alpha$ continues from the zone~$\Omega_\alpha$ to the 
zone~$\Omega_\beta$ (Fig.~\ref{AdjacentLines}).

 We are not aware of any other general results on the geometry of an individual 
stability zone. It can be seen from Fig.~\ref{cos3DmDisc} that the stability zones 
are not necessarily convex. Moreover, the stability zones can be non-simply connected. 
The corresponding example can be constructed as follows.

\begin{example}
Let the level surface of the dispersion law have the shape shown in Fig.~\ref{Neodnosv}. 
Namely, the surface consists of a family of parallel planes connected by thin curved tubes, 
each tube being contained in a small neighborhood of a plane parallel to a fixed 
plane~$\Pi_0$ and being symmetric with respect to it. Then each family of parallel planes 
forming a not too small angle with~$\Pi_0$ cuts all these tubes along closed curves. 
One can see from the constructions of the works~\cite{dynn1,dynn2} that in 
this case all non-closed components of sections of this surface by planes from 
this family will be topologically regular.
 
 Thus, in this case, all directions $\mathbf B$ that are not too close to the normal 
to the~$\Pi_0$ plane will be in the same stability zone. At the same time, it can be shown 
that this stability zone will not cover the entire sphere.

\begin{figure}[t]
\vspace{5mm}
\includegraphics[width=0.7\linewidth]{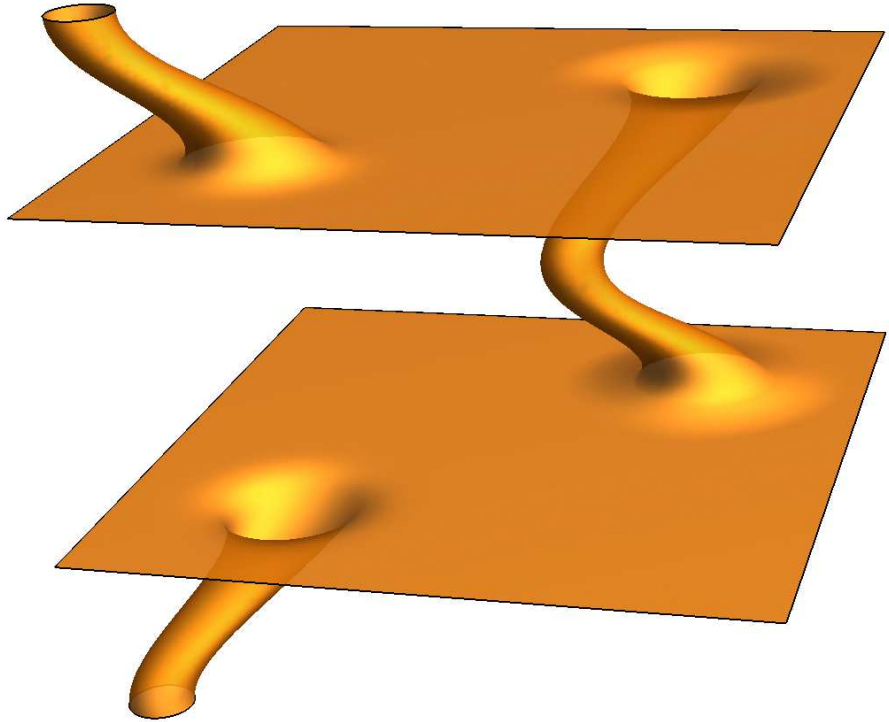}
\caption{Fundamental domain of the surface for which a non-simply connected stability zone 
arises
}
\label{Neodnosv}
\end{figure}
\end{example}

 The functions~$\epsilon_1$ and~$\epsilon_2$ introduced in Lemma~\ref{lem3}, 
generally speaking, need not be continuous on the entire sphere. However, as noted 
in~\cite{dynn3}, they become continuous when restricted to the set of completely 
irrational directions $\mathbf B$, coinciding on this set with the restriction 
of some functions $\widetilde{\epsilon}_{1} (\mathbf B)$,
$\widetilde{\epsilon}_{2} (\mathbf B)$ well defined and continuous on the entire unit 
sphere $\mathbb{S}^{2}$. Inside the stability zones, these functions are also 
uniquely characterized by the fact 
that~$(\widetilde\epsilon_1(\mathbf B),\widetilde\epsilon_2(\mathbf B))$ is the 
largest range of energies~$\epsilon_0$ for which system~\eqref{MFSyst} has stable 
topologically regular trajectories.

 For rational and partially irrational directions $\mathbf B$, the values 
of $\epsilon_{s} (\mathbf B)$ and $\widetilde{\epsilon}_{s} (\mathbf B)$ may not coincide 
due to the presence of unstable topologically regular trajectories at energy levels outside 
the interval~$(\widetilde\epsilon_1(\mathbf B),\widetilde\epsilon_2(\mathbf B))$.
In this case, we always have the inequalities
$$\epsilon_{1} (\mathbf B)\leqslant 
\widetilde{\epsilon}_{1} (\mathbf B) \leqslant 
\widetilde{\epsilon}_{2} (\mathbf B)\leqslant
\epsilon_{2} (\mathbf B) .$$

 Interior points of the stability zones $\Omega_{\alpha}$ are defined by the relation 
$\widetilde{\epsilon}_{1} (\mathbf B) < \widetilde{\epsilon}_{2} (\mathbf B)$. 
The relation $\widetilde{\epsilon}_{1} (\mathbf B) = \widetilde{\epsilon}_{2} (\mathbf B)$ 
takes place at the boundaries of the stability zones, as well as at the accumulation points 
of an infinite number of zones $ \Omega_{\alpha}$ decreasing in size.

 Here we would like to describe the structure of complex angle diagrams in some more detail 
and point out a number of important additional features of such diagrams. Let's start by 
considering the rational directions of the magnetic field, which, in a sense, take a special 
place in an angular diagram.

 As we have already said, rational directions $\mathbf B$ differ from other directions, 
in particular, in that the patterns of trajectories in different planes 
orthogonal to $\mathbf B$ can differ significantly from each other when~$\mathbf B$ is rational. 
All non-singular open  trajectories of the system \eqref{MFSyst}, as well as open trajectories 
in our generalized sense, are periodic in all such planes.

 For rational directions $\mathbf B$, it is typical that at some energy levels in the 
interval $[ \epsilon_{1} (\mathbf B) , \epsilon_{2} (\mathbf B) ]$ periodic complexes of 
stationary points and separatrices are present (Fig. \ref{PeriodicSingular}). To each such 
complex we assign its \emph{rank} which is the dimension of the sublattice 
in~$H_1(\mathbb T^3;\mathbb Z)$ generated by all the separatrix cycles contained in the 
image of this complex under the projection onto the torus~$\mathbb T^3=\mathbb R^3/\mathbb Z^3$.
For example, for the complexes in Fig.~\ref{PeriodicSingular} a)--c) this rank is equal to one, 
and for the complex in Fig.~\ref{PeriodicSingular}~d) it is two. For a fixed rational direction
$\mathbf B$, the presence of such complexes of ranks zero and one is typical, while the 
presence of complexes of rank two requires some additional conditions (certain symmetry or conditions 
of codimension 1). As follows from the results of \cite{dynn1,dynn2,dynn3}, if for a given 
rational direction $\mathbf B$ at least for one energy value
$$\epsilon \in [ \epsilon_{1} (\mathbf B) , \epsilon_{2} (\mathbf B) ] $$
there are only regular trajectories and/or complexes 
of stationary points and separatrices of rank~$\leqslant1$
in all planes orthogonal to $\mathbf B$, then the given direction 
$\mathbf B$ lies inside some stability zone $\Omega_{\alpha} $. It is
natural to call rational directions~$\mathbf B$ of this type \emph{ordinary} rational directions.

Rational directions $\mathbf B$ such that, for any value of 
$\epsilon \in [ \epsilon_{1} (\mathbf B) , \epsilon_{2} (\mathbf B) ]$, there is
a complex of stationary points and separatrices of rank two in at least one plane 
orthogonal to $ \mathbf B$, 
we will call \emph{special} rational directions.

 Special rational directions $\mathbf B$ can appear in different parts of an angle diagram. 
For example, suppose that, for some rational direction $\mathbf B$, there is a periodic complex of separatrices shown 
in Fig.~\ref{PeriodicSingular},d  in one of the 
planes~$\Pi$ orthogonal to $\mathbf B$. The shown complex contains, up to a shift by a reciprocal 
lattice vector, two saddle points in the $\mathbf p$-space connected by separatrices.

 Let us first assume that the gradients of the function $\epsilon (\mathbf p)$ at these 
points are codirectional. Then, for small shifts of the plane keeping its direction, 
the shown complex splits into closed non-singular trajectories of system \eqref{MFSyst}. 
In this case, however, the type of such trajectories (electron or hole) is different for 
shifts in opposite directions.

 It can be shown that the direction $\mathbf B$ lies in this case inside some stability 
zone $\Omega_{\alpha}$, and the corresponding plane~$\Gamma_{\alpha}$ is parallel to $\Pi$.

 The described situation provides an example when a stability zone $\Omega_{\alpha}$ contains 
the direction orthogonal to the corresponding plane $\Gamma_{\alpha}$. In this case, the planes 
orthogonal to $\mathbf B$ may contain periodic trajectories of different directions, periodic 
trajectories of a single direction, or not contain regular periodic trajectories at all. 
The different cases that arise in this situation, in particular, give different regimes of 
behavior of the conductivity tensor in strong magnetic fields (see \cite{UFN}, for example).

 Let us now consider the situation when the gradients $\epsilon (\mathbf p)$ at nonequivalent 
saddle points in Fig.~\ref{PeriodicSingular}, d are opposite to one another. Then, 
for small shifts of the plane~$\Pi$ keeping its direction, the shown complex 
splits into periodic regular trajectories of system \eqref{MFSyst}, which are stable under
small changes of the energy $\epsilon$, as well as under small rotations of the direction 
of $\mathbf B$ around their mean direction. For close partially irrational directions 
$\mathbf B$ obtained as a result of such rotations, the non-degeneracy of the interval 
$[ \epsilon_{1} (\mathbf B) , \epsilon_{2} (\mathbf B) ]$ means that such a direction belongs 
to some stability zone $\Omega_{\alpha}$ or its boundary. It can be seen, therefore, that if 
the original direction $\mathbf B$ does not belong to any stability zone or its boundary, 
it always represents an accumulation point of of stability zones on the angular diagram.

 The above argument is, in fact, of a general nature, which allows us to conclude that any 
special rational direction $\mathbf B$ either belongs to some stability zone (or its boundary), 
or is an accumulation point of stability zones $\Omega_{\alpha}$ on an angle diagram. 
In particular, the above arguments imply the assertion of the theorem~\ref{everywheredense-th} 
that the union of all stability zones $\Omega_{\alpha}$ is everywhere dense on the unit sphere 
$\mathbb{S}^{2} $.

\begin{figure}[t]
\includegraphics[width=\linewidth]{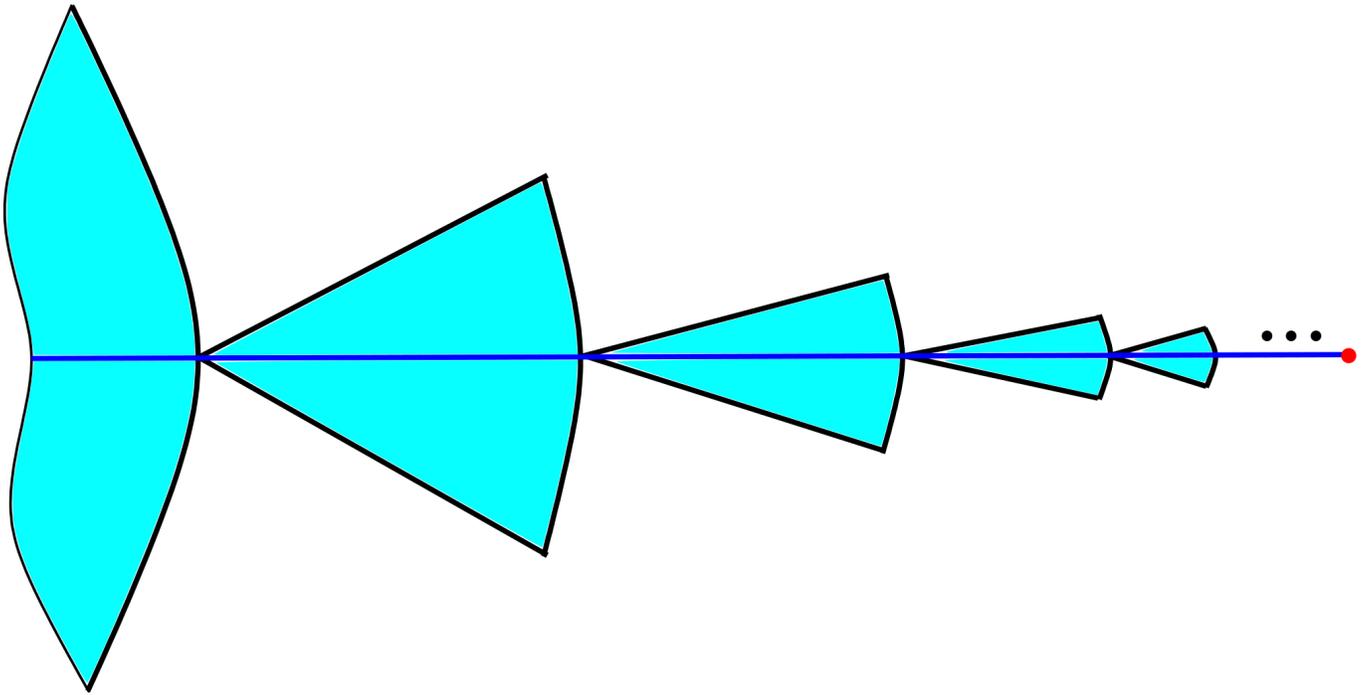} 
\caption{A sequence of stability zones, united by one direction of 
periodic open trajectories, converging to some special rational 
direction $\mathbf B$ (schematically)}
\label{Sequence}
\end{figure}

 In addition to special rational directions $\mathbf B$, the directions $\mathbf B$ 
for which the open trajectories of system \eqref{MFSyst} are chaotic are also accumulation points 
of stability zones. As we noted above, for such a direction of the magnetic 
field~$\mathbf B$ chaotic trajectories present only at one energy level 
$\epsilon_{0} (\mathbf B)$.

\begin{figure}[t]
\begin{tabular}{cc}
\includegraphics[width=0.45\linewidth]{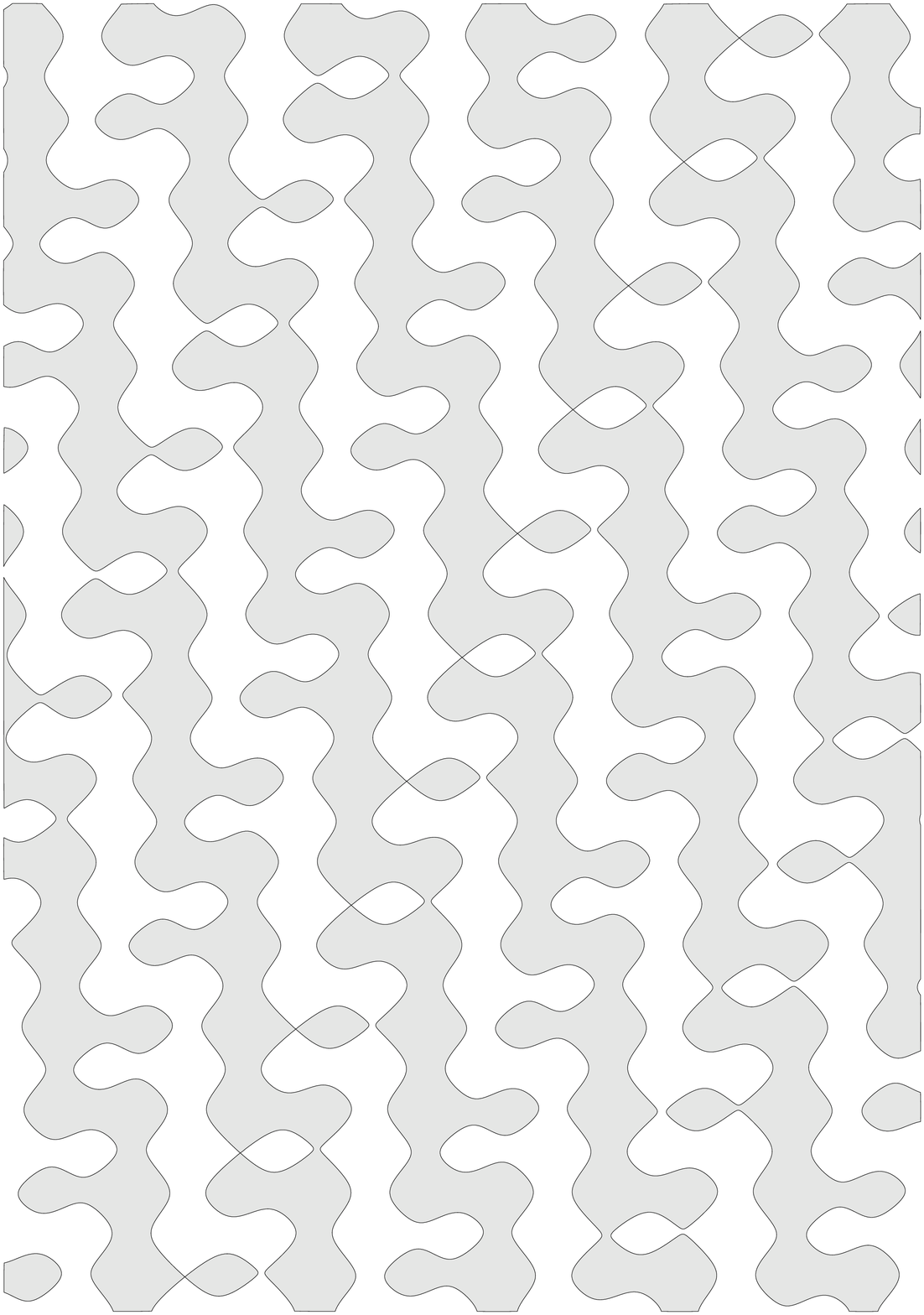}  &
\hspace{5mm}
\includegraphics[width=0.45\linewidth]{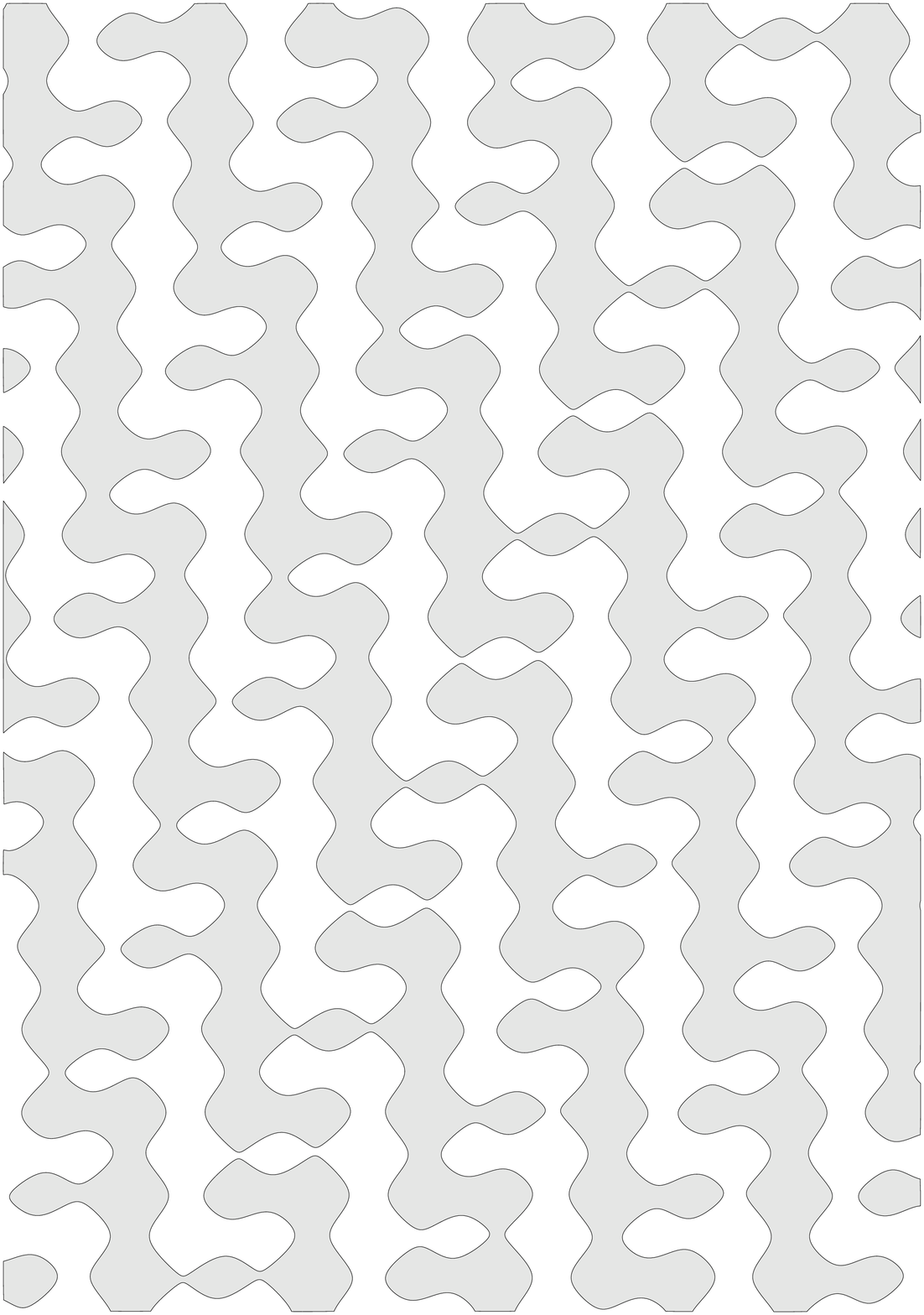}
\end{tabular}
\caption{Examples of topologically regular trajectories and trajectories 
of Tsarev's type in a plane orthogonal to $\mathbf B$}
\label{RegularTsarev}
\end{figure}

 As we also said, the chaotic trajectories of system \eqref{MFSyst} can be divided into 
two main types: Tsarev-type trajectories and Dynnikov-type trajectories. The former can arise 
only for partially irrational directions $\mathbf B$ and always have an asymptotic 
direction in the $\mathbf p$-space (\cite{Tsarev,dynn2}). Unlike regular open trajectories, 
Tsarev-type chaotic trajectories are generally not limited to straight strips of finite width 
in planes orthogonal to $\mathbf B$ (Fig.~\ref{RegularTsarev}). The contribution of Tsarev-type
trajectories to the magnetic conductivity is generally similar to the contribution of 
topologically regular trajectories, although it also has some special features.

\begin{figure}[t]
\includegraphics[width=\linewidth]{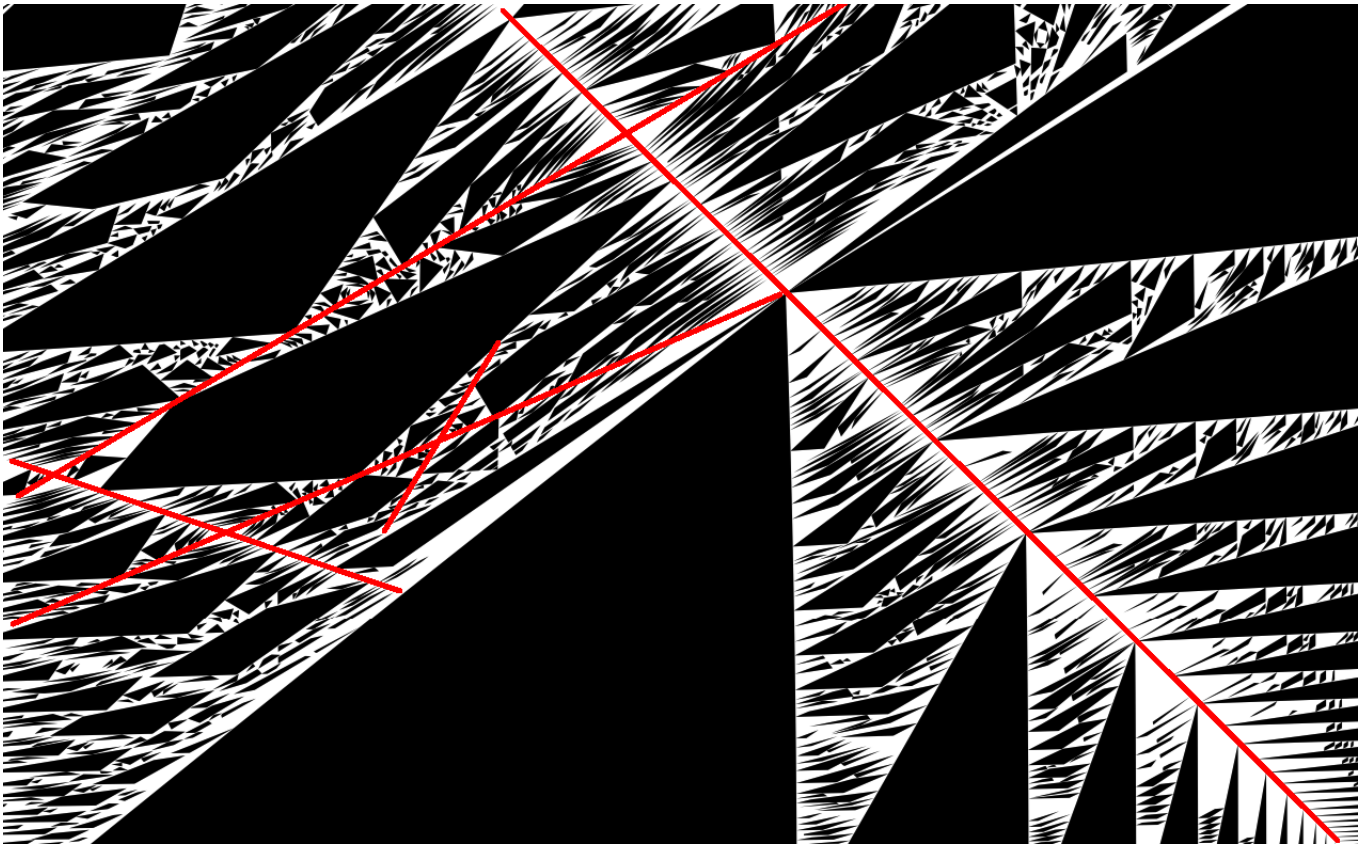} 
\caption{Stability zones and lines of appearance of Tsarev's trajectories 
for a surface $\cos p_{x}+\cos p_{y}+\cos p_{z} =0$}
\label{Diagram}
\end{figure}

 A necessary condition for the presenece of Tsarev-type trajectories 
is the presence on the corresponding level surface 
$$\epsilon (\mathbf p) = \epsilon_{0} $$ 
of separatrix cycles that are nonhomologous to zero in the torus $\mathbb{T}^{3}$. 
As a consequence of this, the corresponding direction $\mathbf B$ must be orthogonal 
to some rational direction in the $\mathbf p$-space, but itself should not be rational.
Thus, directions $\mathbf B$ for which Tsarev-type chaotic trajectories occur 
are naturally combined into families, each of which is contained in a great circular arc
lying in some integral plane of the crystal lattice, and consists of all 
partially irrational points of this arcs. The rational points of these arcs can then be 
boundary points of stability zones, as well as accumulation points of stability zones 
(special rational directions $\mathbf B $), see Fig. \ref{Diagram}.

\vspace{1mm}

\begin{figure}[t]
\includegraphics[width=\linewidth]{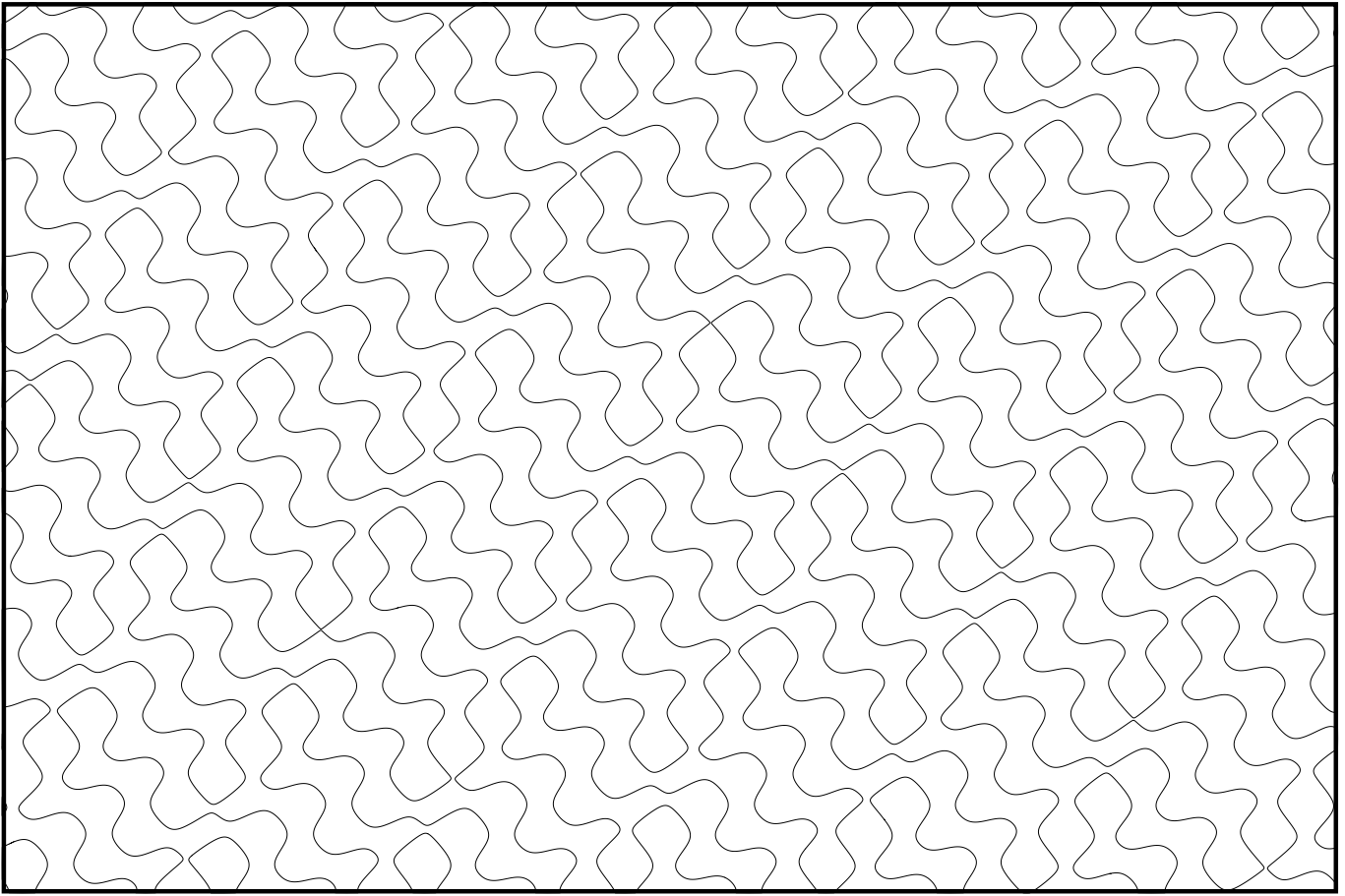} 
\caption{Chaotic Dynnikov-type trajectory lying on the surface
$\cos p_{x}+\cos p_{y}+\cos p_{z} = 0$}
\label{ChaoticTr}
\end{figure}

 A more complex type of chaotic trajectories of system \eqref{MFSyst} are 
Dynnikov-type trajectories (\cite{dynn2}). Trajectories of this type correspond 
to pronounced chaotic dynamics both in planes orthogonal to $\mathbf B$ 
(Fig.~\ref{ChaoticTr}) and on the Fermi surface considered as a compact surface 
in~$\mathbb T^3$. A consequence of such dynamics is the emerging of nontrivial 
regimes of the behavior of conductivity in the presence of such trajectories on the 
Fermi surface (\cite{ZhETF2,TrMian}), among which especially noticeable is the 
suppression of the conductivity in the direction of the magnetic field, as well as 
the presence of fractional powers of $B$ in the asymptotics of the conductivity 
tensor. According to S.P. Novikov's  conjecture~\cite{DynSyst}, the set of directions 
$\mathbf B$ corresponding to chaotic regimes (of any type) for a fixed generic 
dispersion relation has measure zero and the Hausdorff dimension strictly less 
than 2 on the angular diagram. This conjecture is partially confirmed in the following
statement proved in a paper under preparation by
I.Dynnikov, P.Hubert, P.Mercat, O.Paris-Romaskevich, and A.Skripchenko.

\begin{theo}
For a generic dispersion relation $\epsilon$ obeying central symmetry, 
$\epsilon(\mathbf p)=\epsilon(-\mathbf p)$, and such that all its level surfaces viewed 
in~$\mathbb T^ 3$, have genus at most three, the set of directions $\mathbf B$ 
yielding chaotic regimes has measure zero.
\end{theo}

 The study of chaotic trajectories of Dynnikov's type is actively keep going at the 
present time; below we describe some of the most recent results obtained in this area.

\begin{theo}[\cite{dynn4}]
Let the surface~$\epsilon(\mathbf p)=\epsilon_0$ and the magnetic field~$\mathbf B$ 
be such that the trajectories of system~\eqref{MFSyst} are chaotic, and the 
direction $\mathbf B$ is completely irrational. Then in almost all planes orthogonal 
to $\mathbf B$ the number of open trajectories is the same and equals either one, or two,
or is infinite.
\end{theo}

The least difficult is to construct examples for which there is exactly one chaotic trajectory 
in almost every plane orthogonal to $\mathbf B$. For example, all ``self-similar'' examples 
(see~\cite{Skripchenko2}) in the case of a surface of genus three have this property.

 In the papers~\cite{DynnSkrip1,DynnSkrip2}, examples of chaotic regimes in Novikov's
problem are constructed, for which each plane orthogonal to $\mathbf B$ contains an infinite 
number of open trajectories, and these trajectories have an asymptotic direction, but are not 
contained in straight strips of finite width. This effect is related to the absence of unique 
ergodicity of the corresponding foliation on a compact 
level surface~$\epsilon=\epsilon_0$ in~$\mathbb T^3$, and requires a very subtle choice of 
system parameters. Apparently, among the chaotic regimes in Novikov's problem, such a 
situation is not typical.

 No examples are currently known in which almost every plane orthogonal to $\mathbf B$ 
contains exactly two chaotic trajectories. We only know that they do not exist in the case 
of genus three~\cite{dynn4}.

\vspace{1mm}

\begin{figure}[t]
\begin{tabular}{cc}
\includegraphics[width=0.5\linewidth]{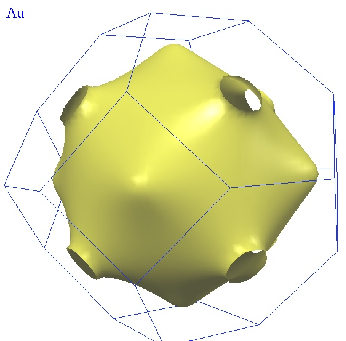}  &
\includegraphics[width=0.45\linewidth]{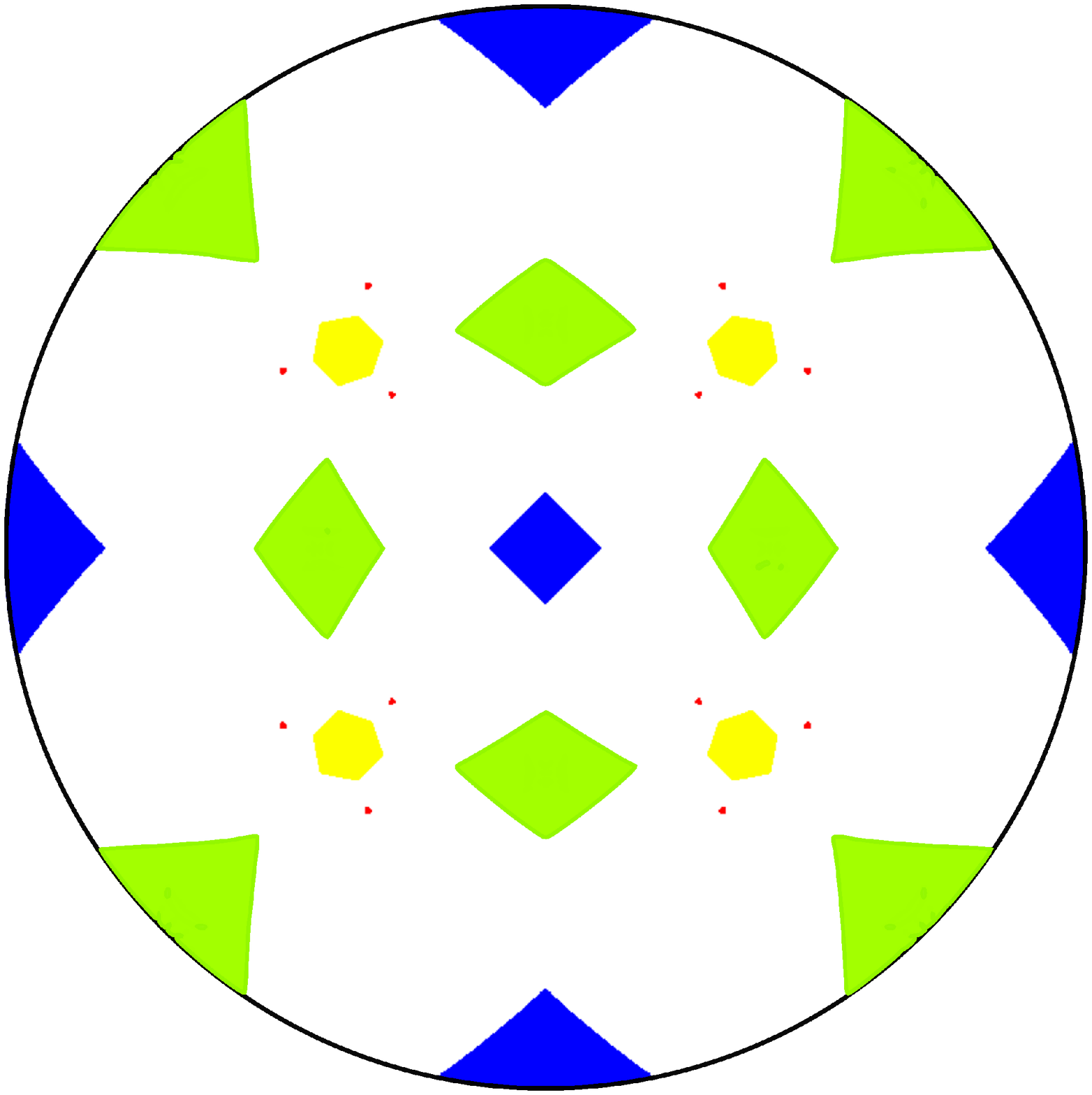}
\end{tabular}
\caption{The Fermi surface of gold and the corresponding exact mathematical 
stability zones in the angular diagram (\cite{DeLeoPhysLettA,DeLeoPhysB})}
\label{Au}
\end{figure}

 When considering galvanomagnetic phenomena in metals, we must take into account only the 
trajectories of system \eqref{MFSyst} that lie at the Fermi level. Accordingly, it is 
natural to introduce angular diagrams showing the presence of open 
trajectories, as well as their type, on the Fermi surface 
$\epsilon (\mathbf p) = \epsilon_{\mathrm F}$ for different directions of the magnetic field.
Such diagrams, of course, are poorer than the diagrams for the total dispersion relation, 
and contain domains consisting of directions $\mathbf B$ for which all trajectories at 
the Fermi level are closed (see e.g. Fig.~\ref{Au}). The stability zones on such diagrams 
are defined as closures of the connected components of the set of directions 
$\mathbf B$ for which system~\eqref{MFSyst} has \emph{stable} topologically regular 
open trajectories at the level~$\epsilon=\epsilon_F$ .

 Each stability zone $\Omega^{*}_{\alpha}$ on such diagrams, as before, is a domain with 
a piecewise smooth boundary on the unit sphere. It is characterized by some values of the 
topological invariants $(m^{1}_{\alpha}, m^{2}_{\alpha}, m^{3}_{\alpha})$ and is a subdomain 
of the corresponding zone $\Omega_{ \alpha}$ defined for the entire dispersion relation.

\begin{figure}[t]
\includegraphics[width=\linewidth]{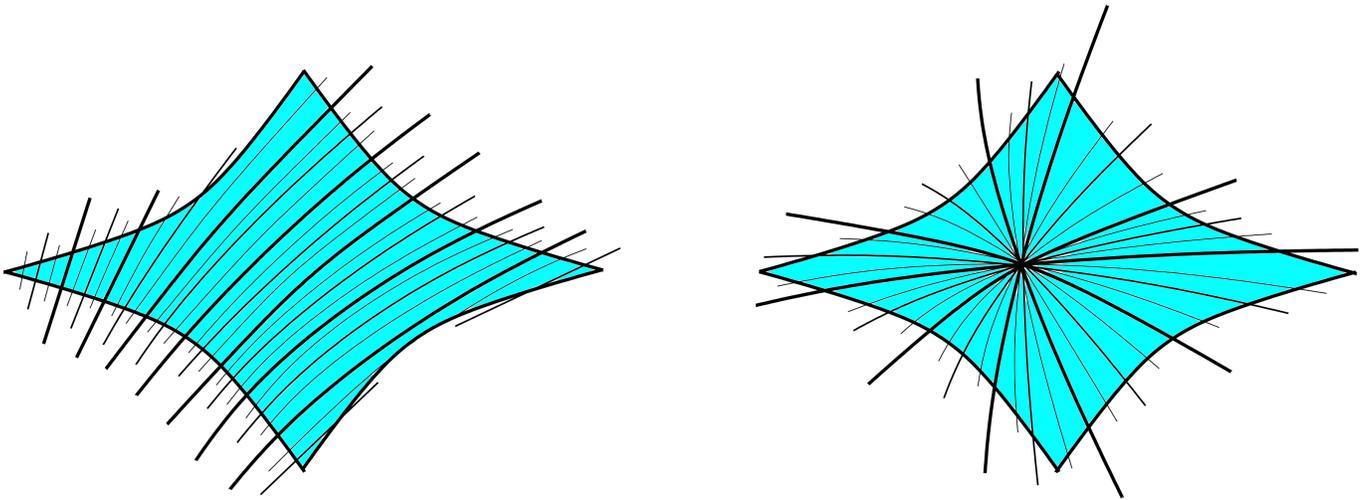} 
\caption{Complete sets of directions $\mathbf B$ inside one stability zone 
$\Omega_{\alpha}$ corresponding to the presence of open trajectories on the Fermi surface}
\label{DirOutStabZone}
\end{figure}

 If, however, we are talking about the complete family of open trajectories on the 
Fermi surface associated with a given stability zone $\Omega_{\alpha}$, then the 
corresponding set of directions $\mathbf B$ goes beyond the zone~$\Omega^*_\alpha$. 
The reason for this is the presence of unstable periodic trajectories for some 
partially irrational directions $\mathbf B$ located near the boundaries of each 
of the zones $\Omega^{*}_{\alpha}$. Such directions form extensions of arcs of partially 
irrational directions $\mathbf B$, corresponding to the presence of stable periodic 
trajectories in the zone $\Omega^{*}_{\alpha}$, beyond the boundaries of this zone 
(Fig.~\ref{DirOutStabZone}). Adjacent segments form an everywhere dense set at the 
boundary of a stability zone, and their length tends to zero as the period of the 
corresponding trajectories increases.

 As was shown in \cite{AnProp}, such a structure leads to rather complex analytic 
properties of the magnetic conductivity tensor both inside the zones $\Omega^{*}_{\alpha}$ 
and near their boundaries. As observed in \cite{CyclRes}, for experimental 
determination of the exact boundaries of the zones $\Omega^{*}_{\alpha}$, it may be better 
to use methods other than direct measurements of conductivity in strong magnetic fields.

\begin{figure}[t]
\includegraphics[width=\linewidth]{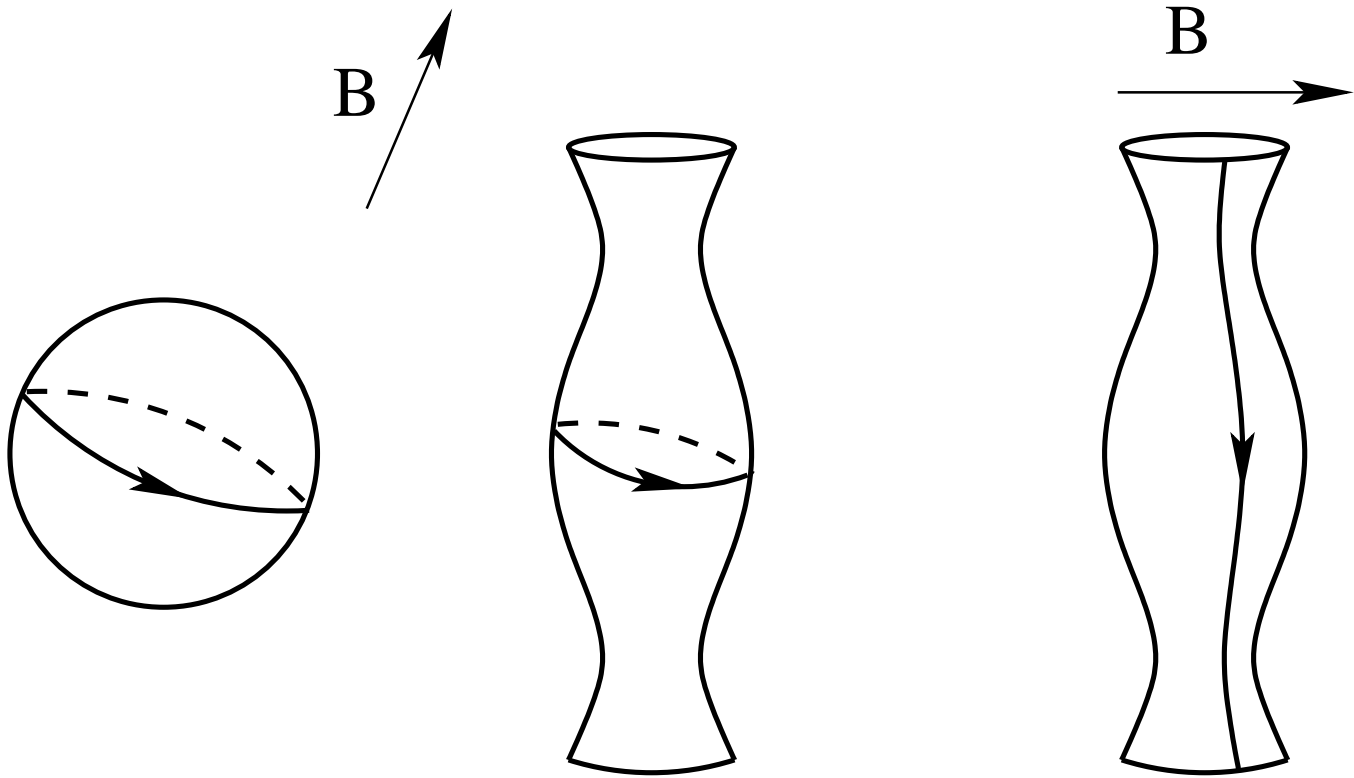} 
\caption{Fermi surfaces corresponding to the simplest angle diagrams}
\label{SimpleSurf}
\end{figure}

 Certainly, angular diagrams for specific Fermi surfaces can be very simple. For example, 
they can entirely consist of directions $\mathbf B$ for which all trajectories of 
system~\eqref{MFSyst} are closed, or they can admit only unstable periodic trajectories 
for some~$\mathbf B$ (Fig.~\ref{SimpleSurf}).

 Certain restrictions on the structure of an angular diagram are imposed by the topology 
of the Fermi surface and its embeddings in the torus~$\mathbb T^3$. An important role here 
is played by such characteristic as the dimension of the image of its one-dimensional 
homology in the one-dimensional homology of the torus $\mathbb{T}^{3}$ 
$$H_{1} \left( S_{\mathrm F} \right) \rightarrow H_ {1} 
\left( \mathbb{T}^{3} \right), $$ 
which we call the \emph{topological rank} of the Fermi surface.
It obviously can take the values 0, 1, 2, and 3. Angular diagrams containing 
stability zones can only emerge for Fermi surfaces of topological rank two or more, 
and more than one stability zone can exist only if the topological rank is three. 
Note that Fermi surfaces of rank 3 must have genus $g \geqslant 3 $.

 As for the presence of directions $\mathbf B$ on the angular diagrams for Fermi surfaces 
that are not included in any of the stability zones, but which are the limit points of the 
union of all stability zones, then such a situation is possible only when the diagram of the 
entire dispersion relation contains an infinite number of stability zones. Such directions 
$\mathbf B$ will be called \emph{singular} for a given Fermi surface. As noted 
in \cite{SecBound}, the presence of singular directions requires that the Fermi 
energy~$\epsilon_{\mathrm F}$ falls into a rather narrow energy interval, determined 
by the dispersion relation $\epsilon (\mathbf p)$.

 Namely, consider some fixed periodic function (dispersion relation) 
$\epsilon (\mathbf p)$, taking values in a certain interval
$$\epsilon_{\rm min} \leqslant 
\epsilon (\mathbf p)\leqslant 
\epsilon_{\rm max}.$$

 It is easy to see that for values of $\epsilon_{\mathrm F}$ close to 
$\epsilon_{\rm min}$ or $\epsilon_{\rm max}$, the Fermi surfaces is very simple, 
and the angular diagrams corresponding to them are trivial (all trajectories of 
the system~\eqref{MFSyst} are closed). One can introduce values
$\epsilon^{\mathcal A}_{1}$,
$\epsilon^{\mathcal A}_{2}$,
$$\epsilon_{\rm min} < 
\epsilon^{\mathcal A}_{1} < 
\epsilon^{\mathcal A}_{2} < 
\epsilon_{\rm max} , $$
such that for the values of the Fermi energy $\epsilon_{\mathrm F}$ lying in the 
interval $( \epsilon^{\mathcal A}_{1} , \epsilon^{\mathcal A}_ {2} )$, the corresponding 
angular diagrams will contain stability zones.

\begin{figure}[t]
\begin{tabular}{cc}
\includegraphics[width=0.45\linewidth]{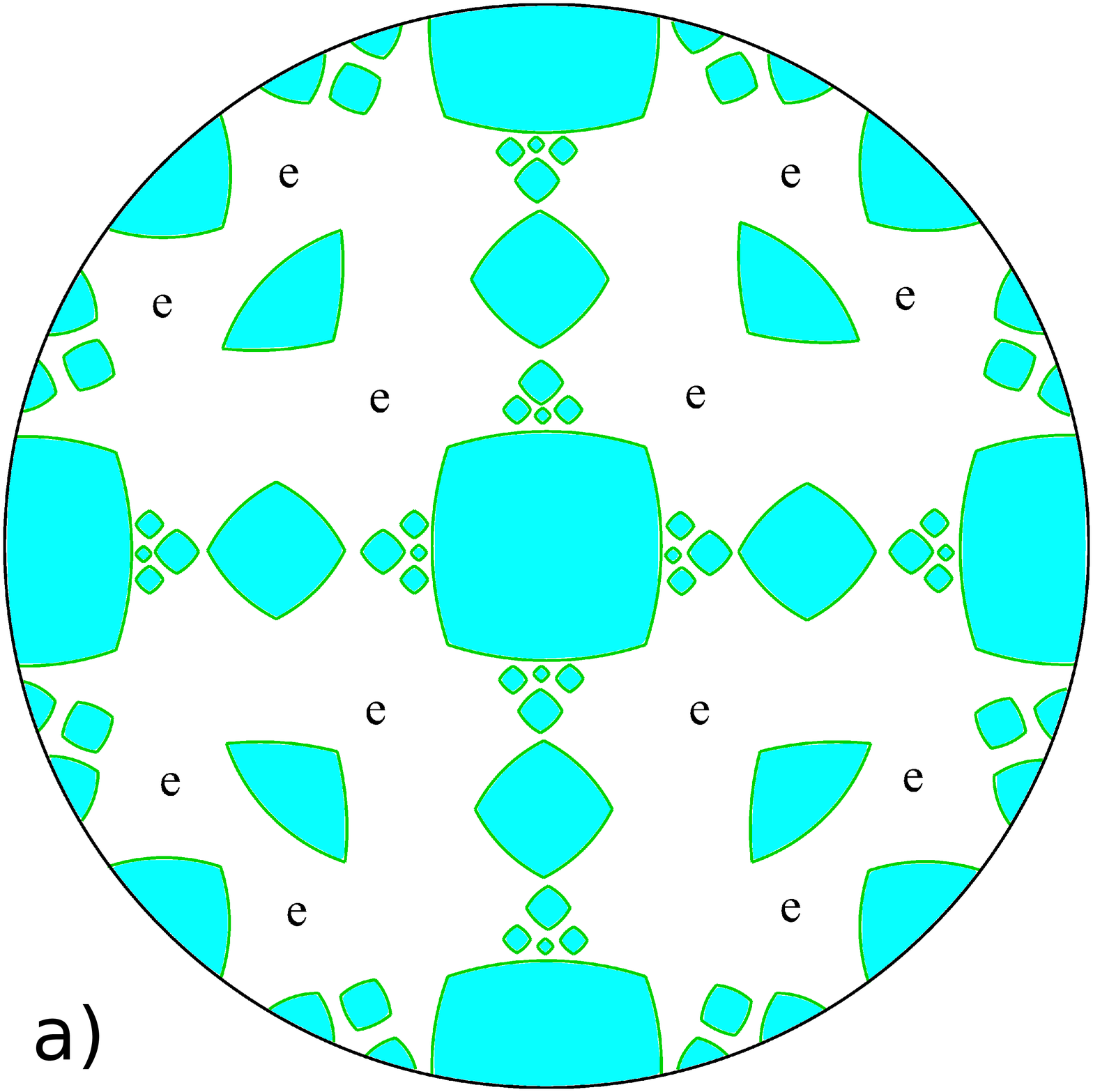}  &
\hspace{5mm} \includegraphics[width=0.45\linewidth]{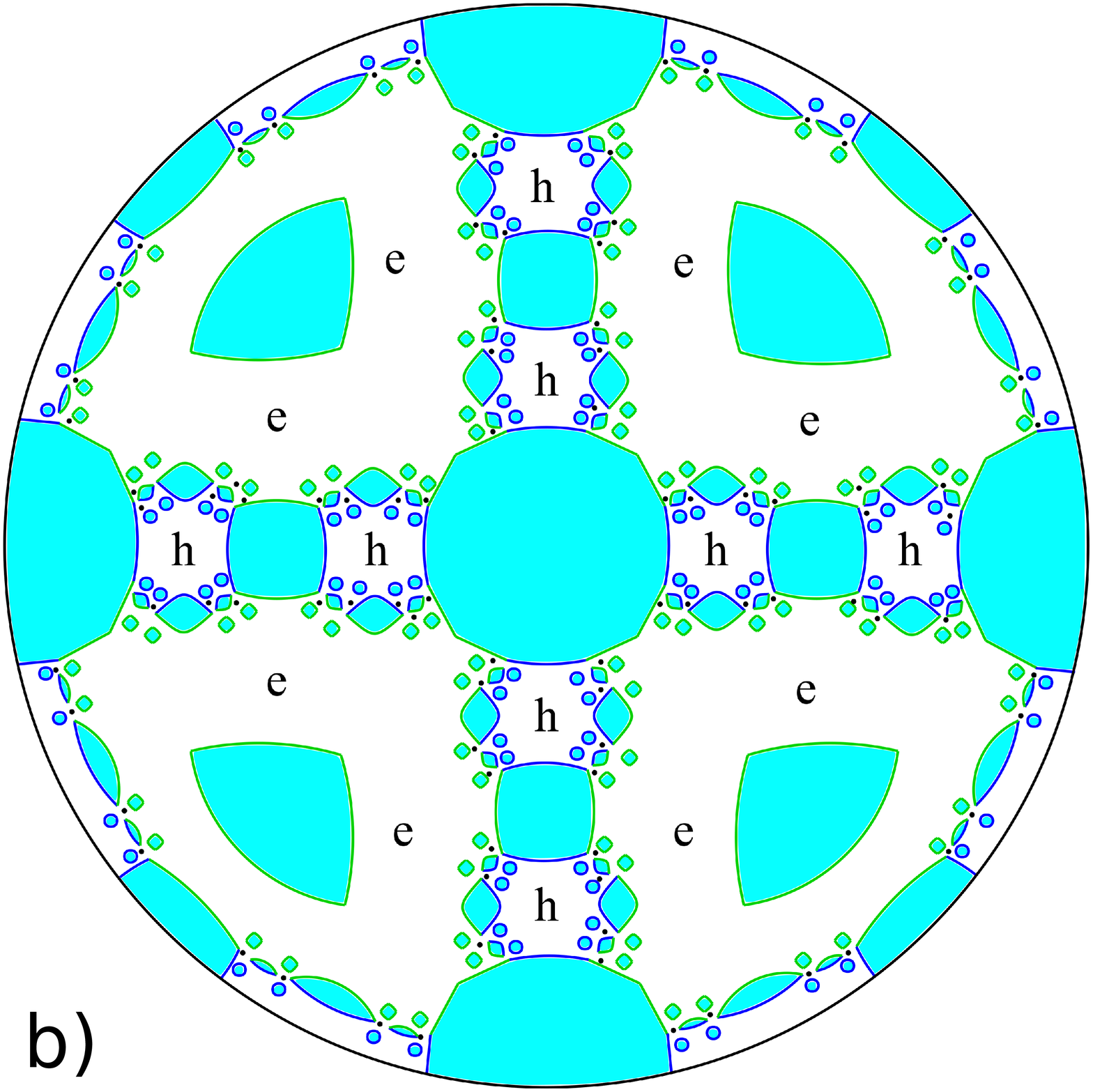}
\end{tabular}
\caption{Complex angle diagrams of type A (a) and type B (b) (schematically). 
The letters e and h denote the type of the Hall conductivity in the domains 
corresponding to the presence of only closed trajectories on the Fermi surface 
(electron and hole, respectively)}
\label{Diagrams}
\end{figure}

 The resulting angle diagrams, in turn, can also be divided into two classes 
(diagrams of type A and diagrams of type B), which differ qualitatively. Namely:

\vspace{1mm}

1) generic diagrams of type A contain only a finite number of stability zones, while 
everywhere in the domain corresponding to the presence of only closed trajectories on the 
Fermi surface, the respective Hall (transverse) conductivity has the same type 
(electron or hole) (Fig. ~ \ref{Diagrams}(a));

\vspace{1mm}

2) generic diagrams of type B contain an infinite number of stability zones, and in the 
domain corresponding to the presence of only closed trajectories on the Fermi surface, 
there are both domains corresponding to electronic Hall conductivity and domains 
corresponding to hole Hall conductivity (Fig.~\ref {Diagrams}(b)).

\vspace{1mm}

 For generic dispersion relations we can define a finite energy interval
$(\epsilon^{\mathcal B}_{1} , \epsilon^{\mathcal B}_{2})$,
$$\epsilon^{\mathcal A}_{1} <
\epsilon^{\mathcal B}_{1} <
\epsilon^{\mathcal B}_{2} <
\epsilon^{\mathcal A}_{2} , $$
such that for all values of $\epsilon_{\mathrm F}$ from this interval the 
corresponding angular diagrams are of type B, while the values of 
$\epsilon_{\mathrm F}$ from the intervals
$( \epsilon^{\mathcal A}_{1} , \epsilon^{\mathcal B}_{1} )$ and
$( \epsilon^{\mathcal B}_{2} , \epsilon^{\mathcal A}_{2} )$
correspond to angle diagrams of type A.

\vspace{1mm}

 Generic diagrams of type A do not contain singular directions ${\bf B}$. 
On the contrary, generic diagrams of type B must contain them. We also note here that, 
since the Fermi levels at which special rational directions $\mathbf B$ can arise 
have measure zero, for generic values $\epsilon_{\mathrm F}$ from the interval
$(\epsilon^{\mathcal B}_{1} , \epsilon^{\mathcal B}_{2})$ singular directions 
$\mathbf B$ will correspond to the presence of chaotic trajectories of Tsarev's
or Dynnikov's type.

 It is shown in \cite{dynn3} that the measure of the set of directions $\mathbf B$ 
corresponding to the presence of chaotic trajectories on a generic Fermi surface is equal 
to zero. According to S.P. Novikov's conjecture (\cite{BullBrazMathSoc,JournStatPhys}), 
the Hausdorff dimension of this set for generic surfaces is strictly less than 1.

 In conclusion, we note here that, apparently, the length of the interval 
$(\epsilon^{\mathcal B}_{1} , \epsilon^{\mathcal B}_{2})$ for actual dispersion relations 
is rather small, which, perhaps, explains the absence, to date, of clear evidence 
of chaotic trajectories of system \eqref{MFSyst} in experiments with real conductors.
It is also possible, of course, that unfamiliarity with trajectories of this type until 
very recently did not allow to interpret certain experimental data appropriately. 
However, we hope that trajectories of this type, as well as the respective behavior of the magnetic 
conductivity, will anyway be discovered in the future for suitable 
classes of conductors among the huge variety of new materials currently being produced.

\section{General Novikov's problem. Setting and results.}

 As we said in the Introduction, the general Novikov's problem is to describe the geometry 
of open level lines of a quasi-periodic function having an arbitrary number of 
quasi-periods on the plane. One of the most natural formulations is to describe the level lines 
of a family of functions obtained by compose a given generic $N$-periodic function~$F$ 
on the space $\mathbb{R}^{N}$ with all possible affine embeddings 
$\iota :\mathbb{R}^{2} \rightarrow \mathbb{R}^{N}$. The global properties of level lines 
of our interest depend primarily on the direction of the plane~$\iota(\mathbb R^2)$, 
which is a point of the Grassmann manifold $G_{N,2}$, and may also depend on the shift 
parameters. The change of affine coordinates in~$\mathbb R^2$ does not play any role for us.

 The most essential result in Novikov's problem for~$N>3$ is the following statement.

\begin{theo}[\cite{NovKvazFunc,DynNov}]
There is an open everywhere dense subset $S \subset C^{\infty} (\mathbb{T}^{4})$ 
of 4-periodic functions $F$, and an open everywhere dense subset $X_{F} \subset G_{4, 2}$ 
depending on $F$ such that for any $\xi \in X_{F}$, any level line of the restriction of the 
function~$F$ onto any two-dimensional plane of direction $\xi$ in~$\mathbb R^4$ 
is contained in a straight strip of finite width. The widths of these strips, as well as 
the diameters of the compact level lines, are bounded above by a constant that depends 
only on the pair~$(F,\xi)$, and any open non-singular level line passes through the 
corresponding strip (Fig.~\ref{StableTr}). The directions of the strips containing open 
level lines are orthogonal to some integer vector~$(m^1,m^2,m^3,m^4)$ which is a locally 
constant function of~$(F,\xi)$.
\end{theo}

 It is easy to see that the situation of stable ``topologically regular'' behavior 
of the level lines of a quasi-periodic function on the plane can occur for any number 
of quasi-periods. For example, it will take place if for the corresponding $N$-periodic 
function in~$\mathbb R^N$ we take a small perturbation of a periodic function depending 
on only one coordinate. Thus, for a family of quasi-periodic functions obtained as 
restrictions of a fixed $N$-periodic function to all possible planes, one can also define 
stability zones, which are open domains in~$G_{N,2}$. It should be noted, however, 
that with an increase in the number of quasiperiods, the shape of topologically regular 
level lines often is getting more complicated, approaching chaotic behavior on finite scales. 
In general, as the number of quasi-periods increases, Novikov's problem approaches more 
and more a problem on random potentials in the two-dimensional plane.

 Below we present a number of topological results related to Novikov's problem with 
an arbitrary number of quasi-periods that generalize the previously known statements 
for the case~$N=3$.

\begin{lemm}
Let $F (z^{1}, \dots, z^{N})$ be an $N$-periodic function with respect to some integer 
lattice in $ \mathbb{R}^{N} $, and let ~$\xi\in G_{N,2}$ and~$c\in\mathbb R$ be such 
that for any two-dimensional affine plane $\Pi$ of direction $\xi$ all level 
lines~$F|_\Pi=c$ are compact. Then the diameters of all these level lines are bounded 
from above by one constant common to all planes of direction $\xi$.
\end{lemm}

 This fact follows from the compactness of the image of the surface~$F=c$ in the 
torus~$\mathbb T^N$: each compact level line (both singular and non-singular)
has a neighborhood of finite diameter such that all other level lines 
$F =c$ in parallel planes that intersect this neighborhood lie entirely in it. 
From such neighborhoods, one can choose a finite family so that their images cover the 
entire image of the surface~$F=c$ in the torus~$\mathbb T^N$.

 Note that in the case~$N=3$ the assumption that all level lines $F=c$ are compact in 
all planes of a given direction is not required (see Lemma \ref{lem1}).

\begin{theo}\label{th3.3}
Let $F (z^{1}, \dots, z^{N})$ be an $N$-periodic function, and let~$\xi\in G_{N,2}$ 
be a fixed direction of two-dimensional planes in~$\mathbb R^N$. Then the set of 
values~$c\in\mathbb R$ such that for some plane~$\Pi$ of direction~$\xi$ the 
level set~$F|_\Pi=c$ has unbounded components forms a closed interval~$ [c_1,c_2]$ or 
consists of a single point~$c_0$.
\end{theo}

 The proof mostly follows the lines of the proof of a similar assertion for the case~$N=3$ 
given in~\cite{dynn2,dynn3}. Note that there is again a difference from the case~$N=3$, 
which consists in the fact that we consider the entire set of planes of one direction 
simultaneously, while in the case of three quasi-periods the statement was true for 
each individual plane of this direction. In the case of more than three quasi-periods, 
we admit a situation in which all 
level lines~$F|_\Pi=c\in[c_1,c_2]$ in some planes~$\Pi$ of direction~$\xi$
are compact, but their diameters are not limited  from above.

 Note, that if~$\xi$ is a completely irrational direction, which is not contained 
in a hyperplane of rational direction, then for any~$c\in[c_1,c_2]$ (or~$c=c_0$) 
and any plane~$\Pi$ of direction~$\xi$, the level set~$F|_\Pi=c$ has either 
unbounded or arbitrarily large closed components.

 Indeed, let there be a plane $\Pi$ of direction $\xi$, in which there are only 
closed level lines $F|_{\Pi} = c$, the size of which is limited by one constant. 
By the size of a trajectory we mean the minimum diameter of the circle $D$ in 
which it can be placed.

 Let's take a plane of direction $\xi$, in which there is an unbounded level line. 
Let us take a segment of this level line such that it does not contain singular 
points and cannot be enclosed in a circle of diameter $D$. There is a small $\delta$ 
such that this segment is preserved for all parallel shifts of this plane in all 
transversal directions by a distance less than $\delta$.

 Consider the corresponding $\delta$ neighborhood $O$ of our plane (which contains 
unbounded components). Integer shifts of the original plane $\Pi$ are everywhere 
dense in $R^N$ for our embedding and must fall in a neighborhood $O$. 
We get a contradiction.

\vspace{1mm}

 Recall once again that we call here unbounded components not only non-singular 
level lines, but also connected complexes containing singular points and separatrices 
connecting them, unbounded in the plane (the same for bounded components). We also 
impose general position conditions for completely irrational embeddings, namely, 
we require that each such complex contains at most a finite number of multiple 
saddles (of finite multiplicity).

\vspace{1mm}

 We would also like to note here that the statements formulated above can also be 
generalized to a more general case, namely, the case of level surfaces of generic 
quasi-periodic functions in $\mathbb{R}^{n}$ with $N$ quasi-periods. The proofs of 
the corresponding assertions here are similar to those given above under the 
imposition of natural conditions of general position.

\vspace{1mm}

 As can also be shown, the situation can be described more precisely in the 
important nongeneric case of the Novikov problem.

\vspace{1mm}

\begin{theo}
Let $F (z^{1}, \dots, z^{N})$ be an $N$-periodic function, and let~$\xi\in G_{N,2}$ 
be a fixed direction of two-dimensional planes containing exactly one, up to a factor, 
non-zero integer vector and not contained in a rational hyperplane.
Then the following is true for all planes~$\Pi$ of 
direction~$\xi$:
\begin{enumerate}
\item
all non-singular open level lines of the function~$F|_\Pi$ have an asymptotic direction, 
which is the same, up to sign, for all planes of the direction $\xi$;
\item
for a fixed~$c$, the diameters of all compact level lines~$F|_\Pi=c$ are bounded from 
above by a single constant common to all planes of the direction $\xi$;
\item
if, for some $c$, there are non-periodic unbounded level 
lines $F|_\Pi=c$ in some plane $\Pi$, then they also exist in any other plane of direction $\xi$.
\end{enumerate}
\end{theo}

 The proof of the first part follows the proof of a similar assertion for~$N=3$ given 
in~\cite[\S6]{dynn2}. Now, we take for~$\gamma$ a path in $\mathbb R^N$ that lies entirely 
in some plane of direction~$\xi$ and such that the end point of the path~$\gamma$ is 
obtained from the initial point by a shift by $\mathbf w$, where~$\mathbf w$ is an irreducible 
nonzero integer vector parallel to $\xi$. Among all such paths, we choose the one that 
has the least number of intersection points with the surface~$F=c$. Next, in the 
orthogonal complement~$\xi^\perp$, we choose a small neighborhood~$W$ of the origin so that 
the shift of $\gamma$ by vectors from~$W$ does not increase the number of intersection 
points with the surface~$F=c$ . The union~$\Gamma$ of all shifts of $\gamma$ by all 
possible vectors of the form $\mathbf u + \mathbf v$, where~$\mathbf u\in W$, 
$\mathbf v\in\mathbb Z^N$, will cut each plane of direction $\xi$ into strips of a finite 
number of shapes, following in a ``quasi-periodic'' order, and in each strip the pattern of level 
lines is periodic. The rest of the argument does not differ from the case~$N=3$.

 The second and third assertions of the theorem are proved using the same construction.
Compact level lines~$F=c$ in the planes of direction $\xi$ do not intersect~$\Gamma$, 
which means that they are contained in strips of limited width in which the pattern of 
intersection lines is invariant under the shift by the vector~$\mathbf w$. This entails 
the existence of a general upper bound for the diameters of these level lines.

 The presence of non-periodic unbounded level lines $F=c$ in any of the planes of 
direction $\xi$ is equivalent to the fact that the intersection of the surface $F=c$ 
with $\Gamma$ is non-empty. This implies the third assertion of the theorem.

 Note that, as in the case of $N=3$, the presence of an asymptotic direction of the 
level lines in the last theorem does not mean that these level lines lie in straight 
strips of finite width.

\end{document}